\documentclass[pdflatex,sn-basic]{sn-jnl}

\usepackage{graphicx}%
\usepackage{multirow}%
\usepackage{amsmath,amssymb,amsfonts}%
\usepackage{amsthm}%
\usepackage{mathrsfs}%
\usepackage[title]{appendix}%
\usepackage{xcolor}%
\usepackage{textcomp}%
\usepackage{manyfoot}%
\usepackage{booktabs}%
\usepackage{algorithm}%
\usepackage{algorithmicx}%
\usepackage{algpseudocode}%
\usepackage{listings}%

\theoremstyle{thmstyleone}%
\theoremstyle{thmstyletwo}%

\theoremstyle{thmstylethree}%

\begin{document}

\title[Article Title]{On ultra-long period ($53.8$\,min) pulsar ASKAP J1935\,+\,2148: coherent radio emission triggered by local superstrong magnetic reconnection}

\author[1]{Zhi-Yao Yang}
\author*[2,3,4]{Cheng-Min Zhang}\email{zhangcm@bao.ac.cn}
\author*[1]{De-Hua Wang}\email{wangdh@gznu.edu.cn}
\author[5,2]{Erbil Gügercinoğlu}
\author[2,6,7]{Xiang-Han Cui}
\author[8]{Jian-Wei Zhang}
\author[1]{Shu Ma}
\author[1]{Yun-Gang Zhou}

\affil[1]{School of Physics and Electronic Science, Guizhou Normal University, Guiyang 550025, China}
\affil[2]{National Astronomical Observatories, Beijing 100101, China}
\affil[3]{School of Physical Sciences, University of Chinese Academy of Sciences, Beijing 100049, China}
\affil[4]{Key Laboratory of Radio Astronomy, Chinese Academy of Sciences, Beijing 100101, China}
\affil[5]{ School of Arts and Sciences, Qingdao Binhai University, Huangdao District, 266555, Qingdao, China}
\affil[6]{School of Astronomy and Space Science, University of Chinese Academy of Sciences, Beijing 100049, China}
\affil[7]{International Centre for Radio Astronomy Research, Curtin Institute of Radio Astronomy, Perth 6102, Australia}
\affil[8]{Department of Astronomy, Key Laboratory of Astroparticle Physics of Yunnan Province, Yunnan University, Kunming 650091, China}
 







\abstract{The eight ultra-long period pulsars (ULPPs) in radio bands have been discovered recently, e.g., ASKAP J1935+2148 with a spin period of 53.8\,min, which are much longer than those of normal pulsars, spanning from 0.016\,s to 23.5\,s, however the origins, spin evolutions and emission mechanisms of these sources are still puzzling. 
We investigate how the ultra-long period of ASKAP J1935+2148 is evolved by the braking of relativistic particle wind, in a time scale of about 0.1 - 1 Myr, from a normal pulsar with local superstrong magnetic fields. In addition, it is noticed that the ULPPs in the period versus period derivative diagram are much below the ``death line", implying their different characteristics from the normal pulsars.
Five sources (including ASKAP J1935+2148) in total eight ULPPs share the rotational energy loss rates to be lower than their respective radio emission luminosities, a phenomenon that can be accounted for by the sustainable radio bursts induced through the reconnection of locally concentrated magnetic field lines.The diversity and complexity of ULPP radio emissions should be closely related to the presence of magnetic reconnection rather than rotational powered discharges in the gaps. Furthermore, it is suggested that the coherent radio emissions of pulsars may have two origins, one from the rotation-powered electric voltage that accounts for the normal pulsar phenomena and the other from the magnetic reconnection-induced continual radio bursts that account for the ULPP observations.}




\maketitle
\section{Introduction} \label{sec:1}
\begin{figure}
    \centering
    \includegraphics[width=0.8\linewidth]{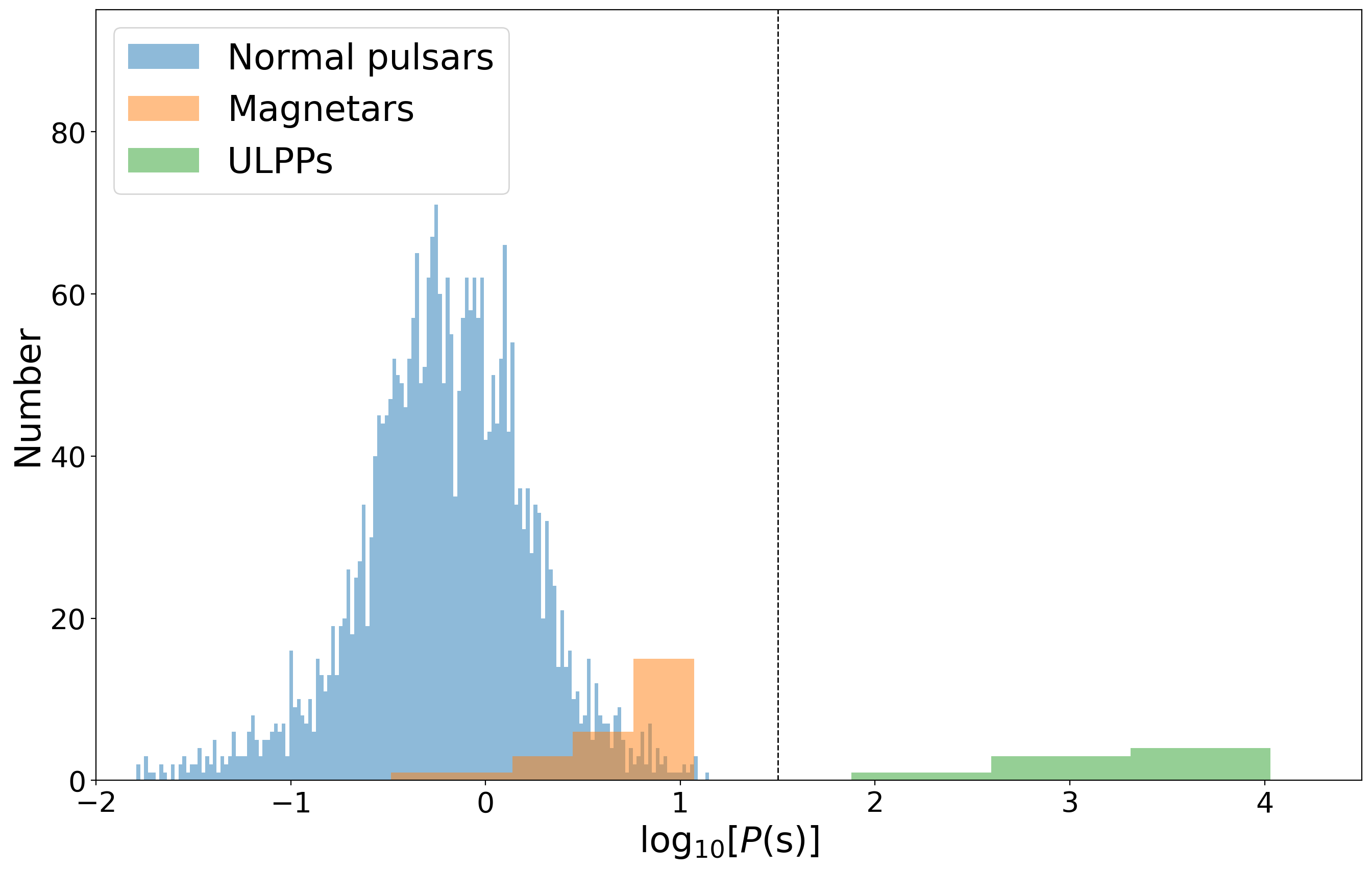}
    \caption{The histogram of spin period distribution for the normal  pulsars, magnetars, and ULPPs (see Table \ref{tab1}). The data are taken from the ATNF Pulsar Catalogue\citep{Manchester05}  and   McGill Online Magnetar Catalogue\citep{Olausen2014}, which are available at  
https://www.atnf.csiro.au/research/pulsar/psrcat/  and 
https://www.physics.mcgill.ca/~pulsar/magnetar/main.html, respectively.}
    \label{fig1}
\end{figure}

Recently, eight radio pulsars with ultra-long periods from $76$\,s to $2.9$\,hr have been discovered as listed in Table \ref{tab1}, e.g., ASKAP J1935\,+\,2148 (hereafter ASKAP J1935) having a spin period of $53.8$\,min. Their distinctive characteristics have called a significant attention, since, as a comparison, the thousands of normal pulsars possess the spin periods in the range $0.016\,\mbox{s}<P<23.5\,\mbox{s}$ \citep{Manchester05}, and their radiation power is mainly derived from the rotational energy, majority of which are located above the `death line' in the so-called  period versus periodic derivatives($P - \dot{P}$) diagram (see Figure \ref{fig3}), with only a few sources located below it. 
However, for magnetars their periods fall in the range $0.3\,\mbox{s}<P<11.78\,\mbox{s}$, and their radiation is  mainly powered by the decay of ultra-strong magnetic field \citep{Ducan1992, Thompson2005,Kaspi17,Esposito2021}, which is manifested in the fact that the energy released by soft gamma-ray repeaters (SGRs) is one to two orders of magnitude higher than their rotational energy loss rate ($\dot{E}$), with the  several exceptions of radio emitting low-magnetic field magnetars \citep{Camilo2006, Camilo2007,Rea10, Levin2012}. Moreover, in the period distribution diagram, there exists  a ``gap" between the normal pulsars and the ultra-long period pulsars (ULPPs), as shown in Figure \ref{fig1}. Therefore, the properties of ULPPs should be  significantly different from those of the normal pulsars and magnetars.

\begin{table}[htbp]
\tabcolsep=0.8pt
\centering
\caption{The physical parameters of ULPPs.}
\begin{tabular}{lccccccccc}
\hline
$\mbox{Source}$ & $P^{a}(\mbox{s})$ & $\dot{P}^{b}(\mbox{s}\cdot \mbox{s}^{-1})$ & $L_{\mbox{R}}^{c}(\mbox{erg}\cdot \mbox{s}^{-1})$ & $B^{d}(\mbox{G})$ & $\dot{E}^{e}(\mbox{erg}\cdot \mbox{s}^{-1})$ & $\mbox{Distance}^{f}(\mbox{kpc})$&$S_{\mbox{R}}^{g}(\mbox{Jy})$& $\mbox{ref.}^{h}$\\
\hline
GLEAM-X J0704-37 & $10496.56$ & $<1.3\times10^{-11}$ & $\sim10^{26}$ &$\sim2.1\times10^{16}$& $4.4\times10^{23}$&$1.5\pm0.5$&$\leq45$& (1)\\
GCRT J1745-3009 & 4620 & - & $\sim10^{30}$ & - & - &$\sim8$&$\sim1$& (2)\\ 
ASKAP J1935 & $3225.31$ & $<1.2\times10^{-10}$ & $\sim10^{29}-10^{30}$ &$\sim2.0\times10^{16}$&$1.4\times10^{26}$&$\sim4.85$&$\sim0.1$ & (3) \\
DART J1832-0911 & $2656.25$ &  $<9.8\times10^{-10}$  &  $\sim10^{31}-10^{32}$  & $\sim5.2\times10^{16}$&$2.1\times10^{27}$&$4.5\pm1.2$&$0.03-20$& (4)\\
GPM J1839-10 &  $1318.20$ & $<3.6\times10^{-13}$ & $\sim10^{28
}$ & $\sim2.2\times10^{15}$&$6.2\times10^{24}$&$5.7\pm2.9$&$0.1-10$ & (5) \\
GLEAM-X J1627 & $1090.8$ & $<1.2\times10^{-9}$ & $10^{28}-10^{31}$ & $\sim1.6\times10^{14}$ &$1.2\times10^{28}$&$1.3\pm0.5$&$5-40$& (6)\\
CHIME J0630+25 & $421.36$ & $<1.6\times10^{-12}$ & $\sim10^{26}$ & $\sim8.5\times10^{14}$ & $\sim8.5\times10^{26}$&$0.17\pm0.08$&$0.5-1.2$& (7) \\
PSR J0901-4046 & $75.88$ & $2.25\times10^{-13}$ & $\sim10^{26}$ & $\sim1.3\times10^{14}$&$2.0\times10^{28}$&$\sim0.40$&$\sim4.1\times10^{-4}$&(8)\\
\hline
\end{tabular}
Notes:
$^{a}$ the spin period;\\
$^{b}$ the spin period derivative;\\
$^{c}$ $L_{\mbox{R}}=7.4\times10^{27}(d/\mbox{kpc})^{2}(S_{\rm R}/\mbox{mJy})$\,erg s$^{-1}$ is the isotropic radio luminosity ; \\
$^{d}$ The characteristic magnetic field strength is calculated by the Magnetic Dipole Radiation (MDR) model $(B=3.2\times10^{19}\sqrt{P\dot{P}})$ \citep{Manchester05, Lorimer12};\\
$^{e}$$\dot{E}=4\pi^{2}I\dot{P}P^{-3}$ is the rotational energy loss rate with $I=10^{45}$\,g cm$^{2}$;\\
$^{f}$ indicate the distance in kpc unit;\\
$^{g}$ indicates the radio flow density in Jy unit\\
$^{h}$:(1) \citep{Hurley-Walker24}; (2) \citep{Hyman05}; (3) \citep{Caleb2024}; (4) \citep{Li2024,WangZiteng2024};  (5) \citep{Hurley22}; (6) \citep{Hurley23}; (7) \citep{Dong2024}; (8) \citep{Caleb22}. 
\label{tab1}
\end{table}


Various ideas for explaining the nature of ULPPs  have been  proposed. One possibility is that these sources could be highly magnetized, isolated white dwarfs \citep{Katz2022, Rea2022} or cataclysmic variables, i.e. white dwarfs in a close binary system with M spectral type companions \citep{Qu2024}, though there is currently insufficient observational evidence to support this hypothesis \citep{Pelisoli2024}. Another suggestion is the interaction of a canonical $\sim10^{12-13}$\,G magnetic field neutron star(NS) with a fallback disk, which allows for the long-period pulsars to appear after $\sim10^{5}$ year of evolution \citep{Gencali2022, Tong2023, Tong23, Yang2024}. Yet next possibility is that the ULPPs are long-term evolution products of high magnetic field magnetars \citep{Beniamini2023} and may have twisted magnetic field structure just above the polar cap due to plastic flow and thermoelectric effect action in their crusts, which could maintain radiative properties and evolutionary processes \citep{Cooper2024}.

In addition to its unusually long spin period, ASKAP J1935 also displayed peculiar pulse emission structure with three distinct states or modes \citep{Caleb2024}: a highly linearly polarized ($\gtrsim90$ per cent), bright mode with pulses lasting 10–50 s, a fainter circularly polarized ($\gtrsim70$ per cent) mode with $\sim100$ ms width pulses, and a quiescent mode with no emission resembling nulling phase of normal pulsars. It has been proposed that the mechanism of generating the unusual radio pulses of the Rotating Radio Transients (RRATs) might also exist in the long-period pulsars \citep{Hurley22, Hurley23, Hyman05}. Due to the fact that the  ULPPs may have unique magnetic field structures, they exhibit the radiation characteristics different from those of normal pulsars \citep{Meszaros92, Michel82, Michel94}. Meanwhile, in high-energy transient events such as Fast Radio Bursts (FRBs), a plausible mechanism for radio emission is the coherent emission generated by the magnetic reconnection process rather than driven by particle acceleration in the gaps sustained via voltage drop by rotation \citep{Bailes2022, Cordes19, Lu18, Zhang2020Natur}.  

Besides, on the formation of the ultra-long spin period of pulsars, it is also suggested that the interaction between pulsars and the surrounding interstellar medium, especially the relativistic particle flows, can lead to the long-term evolution of the pulsar's rotation period, thereby forming the observed ultra-long period characteristics \citep{Rezzolla2018, Ronchi2022, Afonina2024}. Recently, the magnetic dipole radiation + wind (MDRW) model has been proposed to explain the spin-down of the Crab pulsar \citep{Zhang22}, in which the rotational energy loss rate ($\dot{E}$) of pulsars is considered to be sum of the radiation powers of the magnetic dipole radiation ($L_{\rm d} = K_{1}\Omega^{4}$) and the wind ($L_{\rm f} = K_{2}\Omega^{2}$) from the magnetosphere, where $\Omega=2\pi/P$ is the angular speed of the NS with $K_{1}$ and $K_{2}$ being appropriate constants.  

In this work, we consider whether the normal pulsars can evolve into ULPPs through the influence of  particle wind outflows, and explore the emission mechanism of ULPPs by invoking the magnetic reconnection of the local superstrong magnetic field lines of magnetar strength. The structure of the paper is as follows. Section \ref{sec:2} discusses how the ultra-long period of a pulsar is formed, and Section \ref{sec:3} analyzes the emission mechanism of the ULPPs. Subsequently, in Section \ref{sec:4}, we discuss other related implications and concerns.
Finally, we summarize our conclusions in Section \ref{sec:5}.




\section{The process of the period evolution of ULPPs} \label{sec:2}
When exploring the formation of ULPPs such as ASKAP J1935+2148 ($P = 53.8$ min), we can hypothesize that the progenitor stars of such long-period pulsar sources might have evolved from objects similar to the Crab pulsar. 
The age of a pulsar significantly influences its rotational, thermal and magnetic field evolution. The Crab pulsar is the only pulsar whose true age is known (the specific parameters are shown in Table \ref{tab2}  and is regarded as the best astronomical laboratory. 
Therefore, we have selected the Crab pulsar as the reference object for this study. 
Based on the former assumptions, we further examine and analyze the period evolution scenarios under three different models, namely the magnetic dipole radiation (MDR) model, power-law model and MDRW model.

\begin{table}[h]
\centering
\caption{The physical parameters of the Crab pulsar}
    \begin{tabular}{lcc}
    \hline
 Parameter& Value & Ref.\\
 \hline
$I(g\cdot cm^2)$ & $\sim10^{45}$ & \citep{Lorimer12,Haensel07}\\
$R(km)$ & $\sim10$ & \citep{Becker09}\\
$M(M_{\odot})$ & $\sim1.4$ & \citep{Becker09,Miller15}\\
$P(ms)$ & 33.4 & ATNF pulsar Catalogue \\
$\dot{P}(s\cdot s^{-1})$ &  $4.2\times10^{-13}$ &  ATNF pulsar Catalogue \\
$\dot{E}(erg\cdot s^{-1})$ & $4.5\times10^{38}$ &  ATNF pulsar Catalogue \\
 \hline
 \end{tabular}
    \label{tab2}
\end{table}

\subsection{The MDR Model}
The MDR model with a braking index ($n$) equals to 3 serves as the idealized standard model in pulsar astrophysics \citep{Manchester05, Lorimer12, Lyne15, Eksi16, De2022, lyne2022, Araujo2024}. According to MDR, the origin of pulsar radiation results from the conversion of the kinetic energy of a fast rotating magnetized NS. 
As a pulsar rotates, the rotation of the magnetic dipole gives rise to a time varying electric field in the surrounding space, which in turn generates the observed electromagnetic radiation. 
The resulting radiation assumes the form of narrow beams of emission in the vicinity of the magnetic poles. 
As these beams sweep the line of sight, we observe the periodic signals characteristic of pulsars \citep{Manchester77, Lorimer12, lyne2022}. If we assume that the rotation axis is perpendicular to the magnetic dipole moment axis, then the rotational energy loss rate of the pulsar $(\dot{E})$ is equal to the emitted power of the magnetic dipole radiation $(L_{d})$. 
This leads to $\dot{E}=L_{\rm d}$, or the explicit expression given by the following form 
\begin{equation}
	\label{eq1}
	-I\Omega\dot{\Omega}= K_{1}\Omega^{4}\,,
\end{equation}
where $I \sim 10^{45}$\,g$\cdot$cm$^{2}$ is the moment of inertia, 
$\Omega$ and $\dot{\Omega}$ represent the rotational angular speed and its time derivative of the pulsar, respectively, and $K_{1}={(2B^2R^6)}/{(3c^3)}$. Here, the radius $R$ of a NS is about $10$\,km, the speed of light is $c=3 \times 10^{10}$\,cm s$^{-1}$ and the magnetic field $B$ of the pulsar is assumed to be constant throughout of the evolution. 
We can obtain the evolution equation of the spin period under the MDR model as
\begin{equation}
	\label{eq2}
	P(t)=P_{\rm i}[1+(P_{0}/P_{\rm i})^{2}\frac{t}{\tau}]^{\frac{1}{2}}\,,
\end{equation}
where $P_{\rm i}$ is the initial period, $P_{0}$ is the current period and $\tau=P_{0}/2\dot{P}_{0}$ indicates the characteristic age. 
Therefore, 
we obtain the relation between the initial spin $P_{\rm i}$ and the true age of the pulsar as
\begin{equation}
	\label{eq3}
	P_{\rm i}^2=P_{0}^{2}(1-\frac{t}{\tau})\,.
\end{equation}
The current spin period of the Crab pulsar is $P_{0}=33.39$\,ms, and its initial period $P_{i}$ can be derived to be $16.2$\,ms under the MDR model. When evolves to the Hubble age ($T_{\rm{H}}\sim10^{10}$\,yr), the spin period $P(T_{\rm{H}})$ of the Crab Pulsar will reach $94.16$\,s \citep{Zhang22}. This means that the MDR model cannot account for the period of ULPPs.

%
%
%
\subsection{Power-law Model}
If it is assumed that the evolution of the pulsar period obeys a power-law form, then we obtain
\begin{equation}
    \label{eq4}
    P(t)=P_{\rm i}[1+(P_{0}/P_{\rm i})^{n-1}\frac{t}{\tau_{0}}]^\frac{1}{n-1}\,,
\end{equation}
where $n=2.50$ for the Crab pulsar \citep{Lyne15} and  $\tau_{0}=P_{0}/(n-1)\dot{P}_{0}$. The relation between the initial period $P_{\rm i}$ and the true age $t$ can be derived by Equation \ref{eq4} as
\begin{equation}
	\label{eq5}
	P_{\rm i}^{n-1}=P_{0}^{n-1}(1-\frac{t}{\tau})\,.
\end{equation}
In the power-law form, the initial period of the Crab pulsar can be derived to be $P_{\rm i}=18.90$\,ms. According to Equation \ref{eq4}, the spin period $P(t)$ of the Crab pulsar would become $18.30$\,min when evolved to the Hubble age $T_{\rm{H}}\sim10^{10}$\,yr. So even in the power-law form, it is also difficult to explain the period formation of ASKAP J1935.
\subsection{MDRW model}
\begin{figure}
    \centering
    \includegraphics[width=0.5\linewidth]{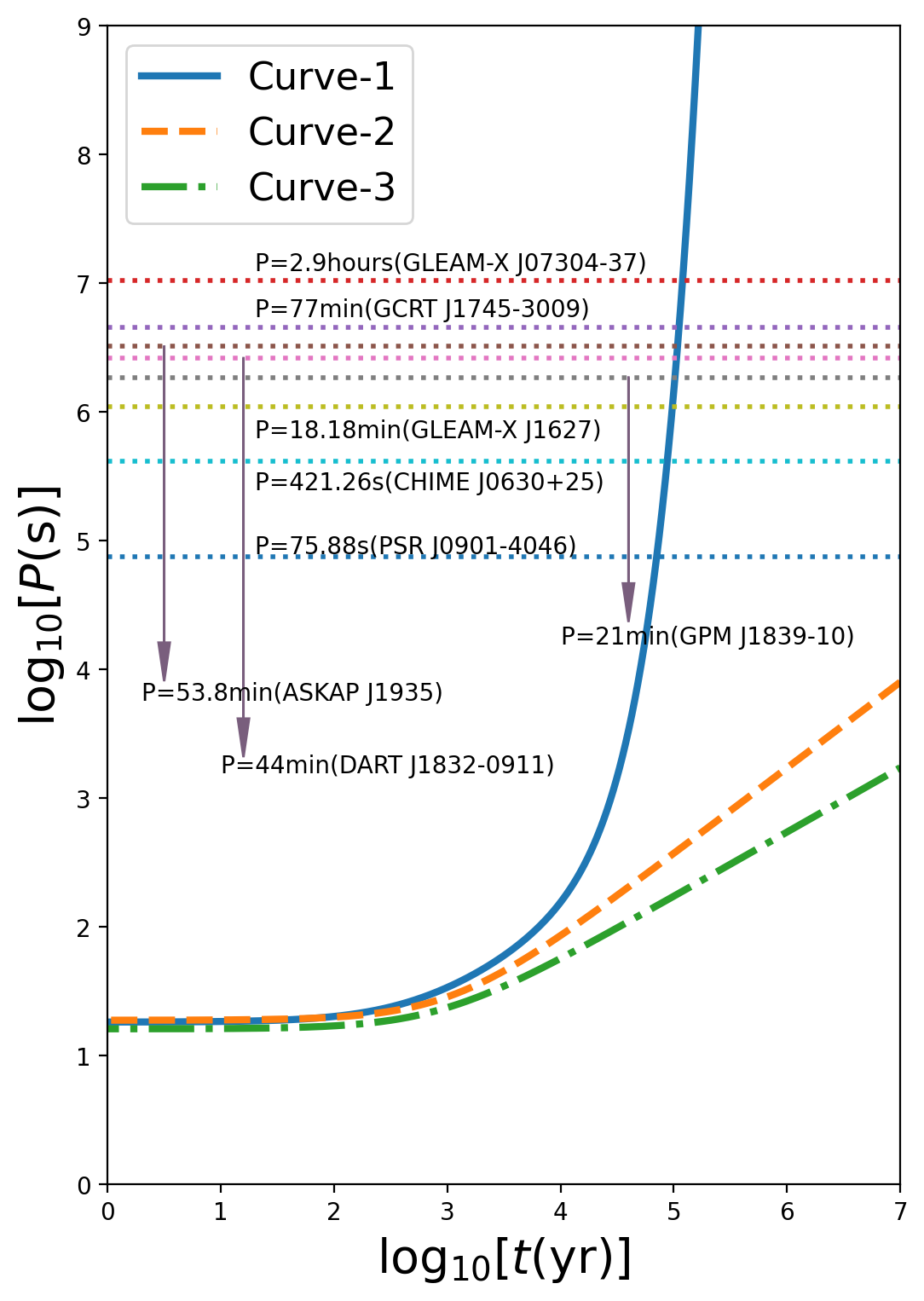}
    \caption{The evolution of the rotation period $(P)$ of the Crab Pulsar with time $(t)$ under the MDRW model, power-law form, and MDR model. Curve-1, Curve-2, and Curve-3 represent the evolutions under MDRW, power-law form, and MDR respectively. The horizontal dotted lines from the top to bottom correspond to the rotation periods of GLEAM-X J0704-37,GCRT J1745-3009, ASKAP J1935, GPM J1839-10, GLEAM-X J1627, CHIME J0630+25 and PSR J0901-4046, respectively (see Table \ref{tab1}).}
    \label{fig2}
\end{figure}

For the Crab Pulsar, it is assumed that its rotational energy loss rate $(\dot{E})$ is equal to the sum of the radiation powers of the magnetic dipole radiation $(L_{\rm d})$ and the particle wind outflow $(L_{\rm f})$,  that is $\dot{E} \equiv L_{\rm d}+L_{\rm f}$, which can be expressed as $-I\Omega\dot{\Omega}=K_{1}\Omega^{4}+K_{2}\Omega^{2}$ \citep{Zhang22}.Among the one undetermined parameters, $K_{2}=({\pi\Phi^2})/({4c})$ is defined by the particle wind flow model \citep{Michel69a},where $\Phi$ is the magnetic flux of the particle flow. The corresponding expression can be simplified to
\begin{equation}
	\label{eq15}
	-\dot{\Omega}=a\Omega^{3} + b\Omega\,,
\end{equation}
where $a={K_{1}}/{I}$ and $b={K_{2}}/{I}$
are the parameters of the magnetic dipole and particle flow components respectively, where the condition $b=0$ or $K_{2}=0$ corresponds to the conventional case of MDR. We adopt the a and b values as suggested by \citet{Zhang22}, i.e., $a\sim2.67\times10^{-16}\,\text{c}\cdot\text{g}\cdot\text{s}$ and $b\sim3.15\times10^{-12}\,\text{c}\cdot\text{g}\cdot\text{s}$, which are inferred by the observational data of the Crab Pulsar.
The corresponding coefficients are $K_{1}=2.67\times10^{29}\,\text{c}\cdot\text{g}\cdot\text{s}$ and $K_{2}=4.21\times10^{33}\,\text{c}\cdot\text{g}\cdot\text{s}$. The relation between the pulsar spin period and its age is obtained from




\begin{equation}
	\label{eq16}
	P=P_{\rm m}\sqrt{(1+\varepsilon_{i})e^{2bt}-1}\,,
\end{equation}
where $P_{m}=57.83$\,ms when $t = 0$, and $\varepsilon_{i}$ is an integration constant, which is defined as the initial value of the fractional ratio ($\varepsilon=\frac{L_f}{L_d}$) between the contributions of the particle flow and magnetic dipole \citep{Lyne15}. In addition, the $\varepsilon_{i}$ value is set to be $\sim 0.1$ as suggested by \citet{Zhang22}, which is inferred by the current spin parameters of the Crab pulsar with the true age of $\sim 960\, \mathrm{yr}$. The analytical solution for the rotational evolution of the Crab pulsar is $P=57.83\sqrt{1.1e^{2bt}-1}$. Therefore, we  find that under the MDRW model the initial period $P_{\rm i}$ of the Crab Pulsar is  $18.29 \, \mathrm{ms}$, and it can evolve to 54 minutes in about 110 kyr with a constant magnetic field of $B=3.3\times10^{12}$\,G.  If we extend the pulsar parameter conditions in  MDRW model, e.g., reducing  the wind braking parameter (b)  by an order of magnitude,  then the time scale for the period of  an ULPP  to  evolve to 54 min is about one million years. Thus, roughly to say, for a Crab-like pulsar evolving to an ULPP of 54 min, it needs about (0.1 - 1) Myr.  This means that the ages of ULPPs should be not too long compared to those of most observed  normal pulsars.  To see the linkage  of normal pulsars and  ULPPs  clearly, the evolutions of the pulsar spin periods based on the  three considered models are shown in Figure \ref{fig2}.


%


\section{Local Superstrong Magnetic Fields in the Polar Cap}\label{sec:3}
\begin{figure}
\centering
    \includegraphics[width=0.8\linewidth]{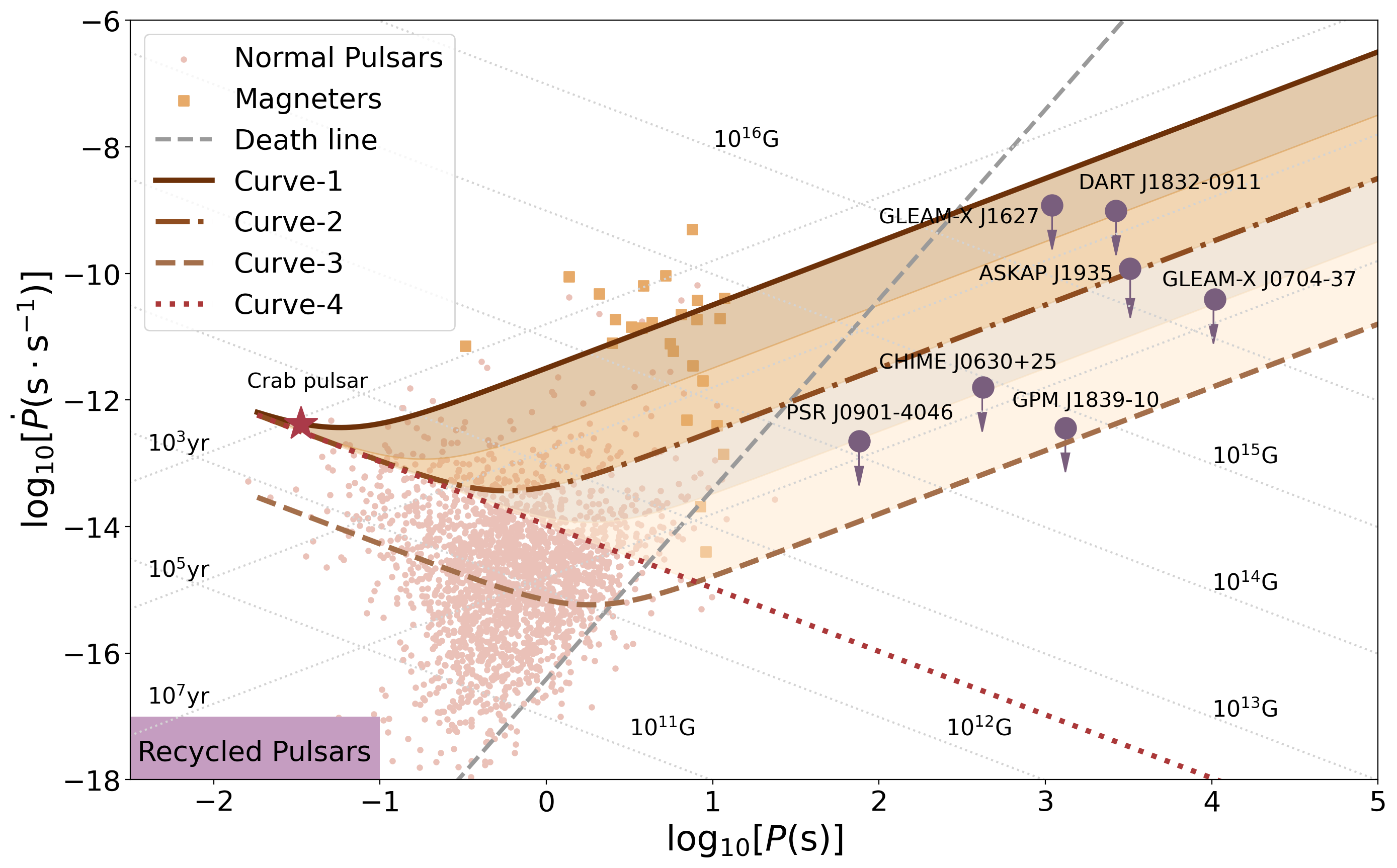}
    \caption{The period and period derivative diagram ($P-\dot{P}$ diagram) of pulsars. The four curves stand for the  different evolution parameters of $a$ and $b$ in MDRW model for the spin period evolution, with the Curve-1 ($a_{1}=2.67\times10^{-16}\,\rm {c\cdot{g}\cdot{s}}$ and $b_{1}=3.15\times10^{-12}\,\rm {c\cdot{g}\cdot{s}}$), the Curve-2 ($a_{2}=a_{1}$ and $b_{2}=0.01b_{1}$), the Curve-3 ($a_{3}=0.1a_{1}$ and $b_{3}=0.001b_{1}$) and the Curve-4 ($a_{4}=0$ and $b_{4}=0.01b_{1}$). The starred sign denotes the Crab pulsar, dots represent the Normal Pulsars, squares represent the magnetars,and the circles represent GLEAM-X J0704-37, ASKAP J1935, GPM J1839-10, DART J1832-0911, GLEAM-X J1627, CHIME J0630+25 and PSR J0901-4046,  respectively. Notes: GCRT J1745-3009 is not marked due to lack of measured $\dot{P}$.}
    \label{fig3}
\end{figure}
The death line of a radio pulsar is defined as a critical condition for the radio emission of pulsar,  
at which the electric potential voltage drop is derived to satisfy the condition of  $(B/10^{12} \rm{G})/P^{2} \geq 0.2$. This can be converted into a relation between the period and the period derivative, as shown in $P-\dot{P}$ diagram (see Figure \ref{fig3}) \citep{Ruderman75, Bhattacharya91}.  In addition, many models have been proposed to extend the criteria of the death line, including the concept of ``Death Valley" proposed by \citet{Chen1993ApJ}, which suggests that the position of the ``death line" depends on the specific configuration of the magnetic field of a NS, e.g., a perfect dipolar magnetic field structure \citep{zhangbing03, Arons00,beskin21}.  However, in reality, the magnetic fields of NSs may exist in the form of some local non-dipolar components ($B_m$) or multipolar components \citep{LaiApJ, Asseo2002, Spitkovsky02, Becker09, Geppert2013, Igoshev2021}. 
The radio luminosity of normal pulsars, thought to be rotation powered NSs, are much less than their energy loss rates \citep{lyne2022, Lorimer12}. However, unlike the radio pulsars,   the radio luminosities of ULPPs are usually higher than their energy loss rates, and these ULPP sources  are located in the death-valley, much below the death-line in $P-\dot{P}$ diagram. Therefore, we suspect that ULPPs may exist with local non-dipolar magnetic fields stronger than the dipole field strength \citep{Petri15, Petri19}, which may account for the observed emission properties of ULPPs. Thus, in the following we investigate the energy loss rates of ULPPs, and their radio emissions induced by reconnection of local superstrong magnetic field. To this end, we take ASKAP 1935 as an example \citep{Caleb2024}. Its rotational kinetic energy $E$ is
\begin{equation}
	\label{eq17}
	E=\frac{1}{2}I\Omega^{2}\sim1.90\times10^{39} \mbox{erg},
\end{equation}
and the rotational energy loss rate $\dot{E}$ is
\begin{equation}
	\label{eq18}
	\dot{E}=-I\Omega\dot{\Omega}=I\frac{4\pi^{2}}{P^{3}}\dot{P}\sim1.67\times10^{26}\, \mbox{erg}\cdot \mbox{s}^{-1}.
\end{equation}
Based on the inferred surface magnetic field of the pulsar ASKAP J1935 as $B\sim10^{16}G$ is a characteristic magnetic field but not a real one, and its local magnetic energy of the NS is
\begin{equation}
	\label{eq19}
	E_{\rm B}=(B^{2}/8\pi)V\sim1.67\times10^{47}(\frac{B}{10^{16} \mbox{G}})^{2}\,(\frac{V}{10^{16}{\rm cm^{3}}})\,\mbox{erg}.
\end{equation}
where 
$V$ represents the volume of the strong magnetic field region around the polar cap of NS. 
Neglecting the distortions due to the magnetic field, the volume of the NS can be approximated by that of a sphere $V=(4\pi R^{3})/(3)$.
From the consideration of local superstrong magnetic fields in the polar cap, its numerical value is $4.19\times10^{16}$\,cm$^{3}$. 
From equations \ref{eq17} and \ref{eq19}, it can be seen that $E_{\rm B}>E$, which means that when the magnetic energy dominates, the radiation from the NSs mainly originates from the energy released by the local super-strong magnetic fields \citep{Bhattacharya1996,petri2020, Pons2019}. The radiation of the ULPP is mainly driven by the magnetic energy dissipation rather than the rotational kinetic energy, via processes like magnetic reconnection. 
Based on equation \ref{eq19}, the magnetic energy release rate ($L_{\rm X}$) can be expressed as
\begin{equation}
	\label{eq20}
	L_{\rm X}=\frac{dE_{B}}{dt}=\frac{BV}{4\pi}\frac{dB}{dt}.
\end{equation}
It is known that the magnetic field of an NS will decay over time, and the decay rate depends on the temperature and electrical conductivity of the material \citep{Shapiro83}. The evolution equation for the crustal magnetic field is \citep{Aguilera2008}
\begin{equation}
	\label{eq21}
	\frac{d{B}}{dt}=-\frac{B}{\tau_{\rm Ohm}}-\frac{B}{B_{\rm i}}\frac{B}{\tau_{\rm Hall}},
\end{equation}
where $\tau_{\rm Ohm}$ is the Ohmic dissipation time-scale given by \citep{Haensel1990, Muslimov1996}
\begin{equation}
	\label{eq22}
	\tau_{\rm Ohm}=\frac{4\pi\sigma L^{2}}{c^2}\approx4.4\mbox{Myr}\left(\frac{\sigma}{10^{24}\,\mbox{s}^{-1}}\right)\left(\frac{L}{1\,\mbox{km}}\right)^{2},
\end{equation}
and the Hall drift time-scale $\tau_{\rm Hall}$ is \citep{Goldreich1992, Gourgouliatos2015}
\begin{equation}
  \tau_{\rm Hall}=\frac{4\pi e n_{\rm e} L^{2}}{c B}\approx6.4\mbox{kyr}\left(\frac{n_{\rm e}}{10^{33}\,\mbox{cm}^{-3}}\right)\left(\frac{L}{1\,\mbox{km}}\right)^{2}\left(\frac{B_{\rm i}}{10^{13}\,\mbox{G}}\right)^{-1}.
\end{equation}
In the above expressions, $B_{\rm i}$ represents the initial magnetic field intensity, $\sigma$ is the electrical conductivity, $L$ is the thickness of the crustal layer that sustains the electric currents, $n_{\rm e}$ and $e$ are the number density and charge of electrons, respectively. The solution to Equation \ref{eq21} is \citep{Aguilera2008} 
\begin{equation}
    B(t)=B_{\rm i}\frac{\exp\left(-t/\tau_{\rm Ohm}\right)}{1+\left(\tau_{\rm Ohm}/\tau_{\rm Hall}\right)\left[1-\exp\left(-t/\tau_{\rm Ohm}\right)\right]}.
    \label{Bte}
\end{equation}
The Hall drift will facilitate the formation of magnetic spots, i.e. small-scale multipolar magnetic field structures at the surface of NSs stronger than the dipole field by draining the toroidal magnetic energy residing in the deeper layers of the crust on timescales of $\gtrsim10^{4}$\,yr \citep{Geppert2014}. In the late stages of the evolution when $t\gtrsim\tau_{\rm Ohm}$, Equation \ref{Bte} approximates to $B\cong B_{\rm i}\exp\left(-t/\tau_{\rm Ohm}\right)$  Based on the above arguments, Equation \ref{eq20} can be simplified to
\begin{equation}
	\label{eq23}
	L_{\rm X}=\frac{1}{2}\frac{E_{\rm B}}{t_{\rm d}},
\end{equation}
where $t_{\rm d}=\mbox{max}[\tau_{\rm Ohm},\tau_{\rm Hall}]$. From Equation \ref{eq23}, it can be inferred that the rate of magnetic energy release is approximately equal to $3.64\times10^{34}$\,erg$\cdot$s$^{-1}$.  Therefore, the local magnetic field evolves over $t_{d}$, that is, the local magnetic field behavior is
\begin{equation}
        \label{eq24}
	B_{*}\sim B_{\rm i}e^{-t/t_{\rm d}}\sim3.68\times10^{15} \mbox{G},
\end{equation}
where $B_{*}$ represents the intensity of the present-day local magnetic field after Ohmic dissipation, and the initial magnetic field intensity $B_i$ is set as the typical value of $\sim 10^{16} {\rm G}$. So, the local magnetic energy is $E_{\rm B}\sim2.26\times10^{44}(B/(10^{15} {\rm G}))^{2}(V/10^{16} {\rm cm^{3}})$\, erg. 
Meanwhile, the magnetic energy release rate is $4.92\times10^{33}$\,erg$\,\cdot$s$^{-1}$. Generally speaking, there is an approximate proportional relationship of $\eta = 10^{-4}-10^{-5}$ between the X-ray luminosity and the radio luminosity \citep{Gudel2008,Fender2000}, that is
\begin{equation}
	\label{eq25}
L_{R}=10^{-4}L_{X}\sim4.92\times10^{29}\,\mbox{erg}\cdot\,\mbox{s}^{-1}.
\end{equation}
In this situation, the radio luminosity induced by magnetic reconnection is compatible with the observational upper limit of $4\times10^{30}$ erg s$^{-1}$ \citep{Caleb2024}. Note that in terms of a magnetar scenario, it is difficult to reconcile a high required dipolar magnetic field to slow down the source with the low implied X-ray luminosity upper limits, which will be easily exceeded by the decay of the magnetic field at mature ages \citep{Suvorov2023}. When the magnetic energy is released, it will cause the high-energy photons to generate electron-positron pairs in the local superstrong magnetic zone. These electrons will be accelerated along the curved magnetic field lines \citep{Giraud2020}, to produce the curvature radiation until they emit radio signals, which is similar to the cascade process scheme for the radio pulsar by Ruderman and Sutherland \citep{Ruderman75}. A similar process, called the tornado-lightening mechanism, in which the discharge of local magnetic field lines and the resulting break-down of the gap above the polar cap  was invoked by \citet{Kontorovich2010} in order to explain the circular polarization and giant pulses seen from normal pulsars.  We stress here that the emissions of normal radio pulsars should be involved in the high electric voltage which is rotation-powered, while the accelerating electric voltage of ULPPs is generated by the  reconnection of local superstrong magnetic field lines through the tornado-lightening mechanism \citep{Parker1957, Melrose1980}. 

\begin{figure}
	\centering
	\includegraphics[width=0.8\textwidth]{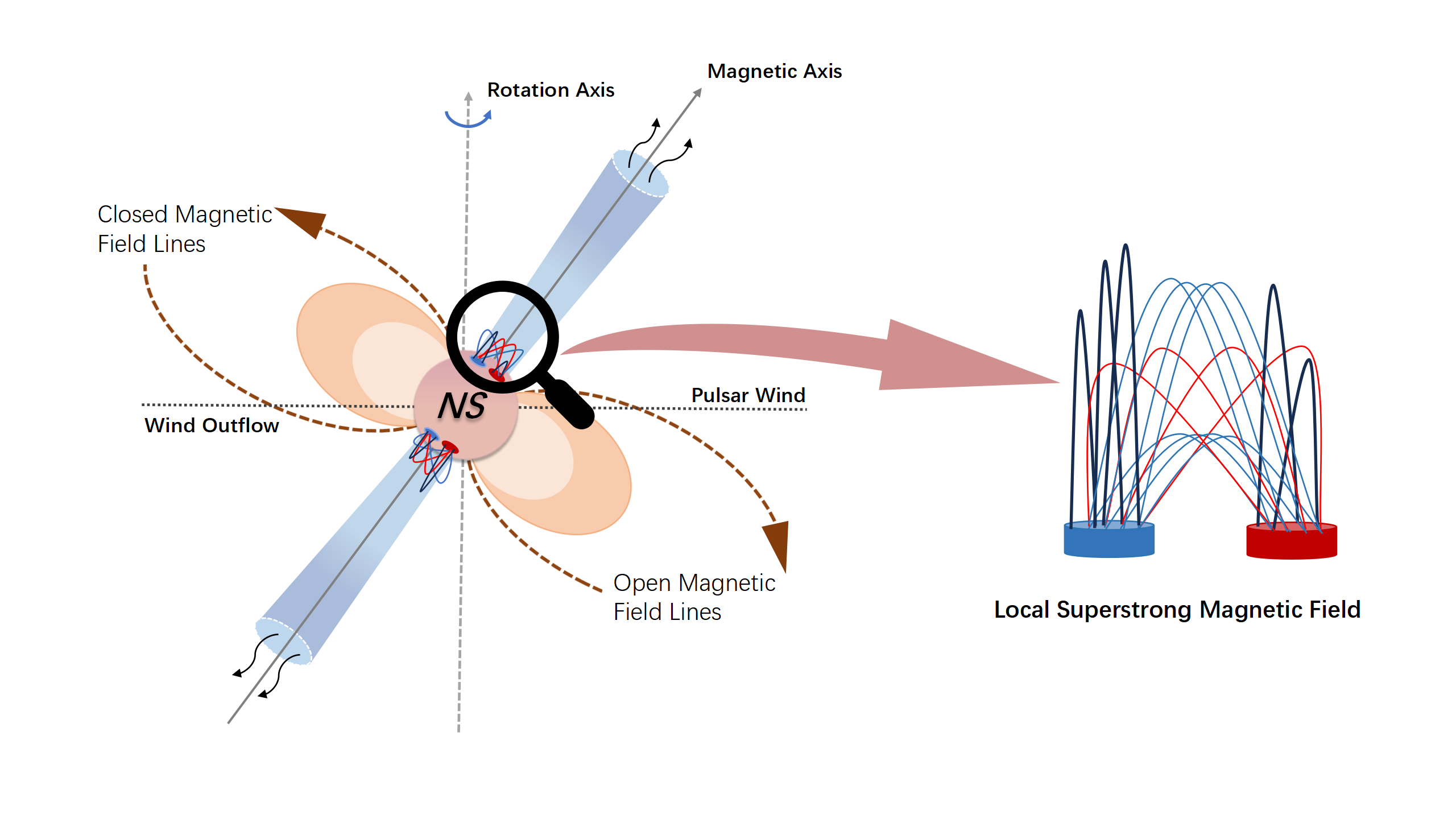}
	\caption{The illustration of NS magnetic field structure, where the local superstrong magnetic in the  polar cap region is shown as zoomed in.}
	\label{fig4}
\end{figure}





\section{Discussion}\label{sec:4}
We initiate the discussion on several pertinent implications of the model, as well as encompassing preliminary considerations of the model assumptions. Furthermore, we delve into the issues of the origin and emission of ULPPs, together with their relations to RRATs, radio magnetars, and normal pulsars.

\subsection{The origin of the ULPP  spin period}

From our model, the distribution of the ULPP periods should be continually filled-in the region from a dozen of seconds to 54 minutes and longer in $P-\dot{P}$ diagram.
However, from the observations, only 8 ULPPs (one of them has no evaluation of the period derivative) have been reported to date, with the period range of  $\sim 1$\,min - $2.9$\,hr. So,    in their period distribution there exists a ``valley" or ``gap" from 10 seconds to 18 minutes. What makes this shortage of samples?  We think that there may exist the selection effects in the ULPP observations, since the radio luminosity of ULPP is in general weak and a long integral time is needed. 
The other possibility may be that the ULPPs are different types of compact objects that are born with slower periods than those of normal pulsars, since their local magnetar-like fields are firstly buried under the crust at birth and diffused out of the NS surface by Ohmic dissipation in about several hundred kyrs while their periods have evolved to over ten minutes via the effective particle wind braking. 
For the situation of the observational selection effects, the astronomers should find more ULPPs with around ten minutes in the future by improving the instrumentation and software; otherwise, for the case of the buried 
local magnetar-like field rebirth, the astronomers may find lots of the ``transient" ULPPs with periods from about one minute to ten minutes.  

\subsection{Local superstrong magnetic field of ULPP} 
Our model suggests that an ULPP has the two components of magnetic fields:  
a dipolar field like the Crab-like pulsar and local magnetar-like multiple fields in the magnetic polar cap region.  
The reconnection of the local superstrong magnetic field is proposed for triggering the coherent radio emissions for the ULPPs, which  should be different from the coherent radio emissions induced by the rotations of the normal pulsars. In detail, near the polar cap of a pulsar, there may exist local regions with concentrated strong magnetic fields such that complex high-order multiple magnetic fields can be present at the surface in the vicinity of the magnetic polar cap \citep{philippov2022}. 
The Hall drift assists magnetic field evolution may play a prominent role in providing and survival of such local strong magnetic field regions at the surface for ages $10^{5}$\,yr. These characteristics may play a crucial role in the emission of the observed pulse signals. Meanwhile, the magnetic field intensity and structure within these regions are different from those of the overall dipolar field. 
These components may change the structure of the magnetic field, leading to complex particle trajectories and radiation patterns \citep{Igoshev2021}. As shown in Figure \ref{fig4}, in this complex environment, two magnetic field lines with opposite polarities in the multiple magnetic field will approach and intertwine in the plasma. Due to the instability of the plasma or changes in local external conditions, the magnetic field lines may tear and reconnect at new positions, forming a new magnetic field topology \citep{Melrose1997, Melrose16, Uzdensky2011, zhang2023}. This process will release a large amount of magnetic energy and simultaneously accelerate the surrounding electrons and ions, generating the high-energy particle streams, which can produce various types of radiation \citep{Uzdensky2016, Schopper1998, De2000}. Additionally, magnetic reconnection may also generate RRATs \citep{MOST2023}. In long-period pulsars, even when the magnetic energy is greater than the rotational energy, that is when the large-scale magnetic field dominates, it does not affect the existence of high-order multiple magnetic fields near the polar cap and the generation of radiation therefrom. This introduces a new proposal to explain the causes of radiation in long-period pulsars. Of course, the radiation of long-period pulsars may be related to the combined effect of the dipolar magnetic field and the local strong multiple magnetic fields. This may account for the ephemeral and distinct emission modes seen from ULPPs, and the subject of which will be further explored in a subsequent study.

\subsection{On the gap between the ULPPs and normal pulsars}
\begin{figure}
    \centering
    \includegraphics[width=0.8\linewidth]{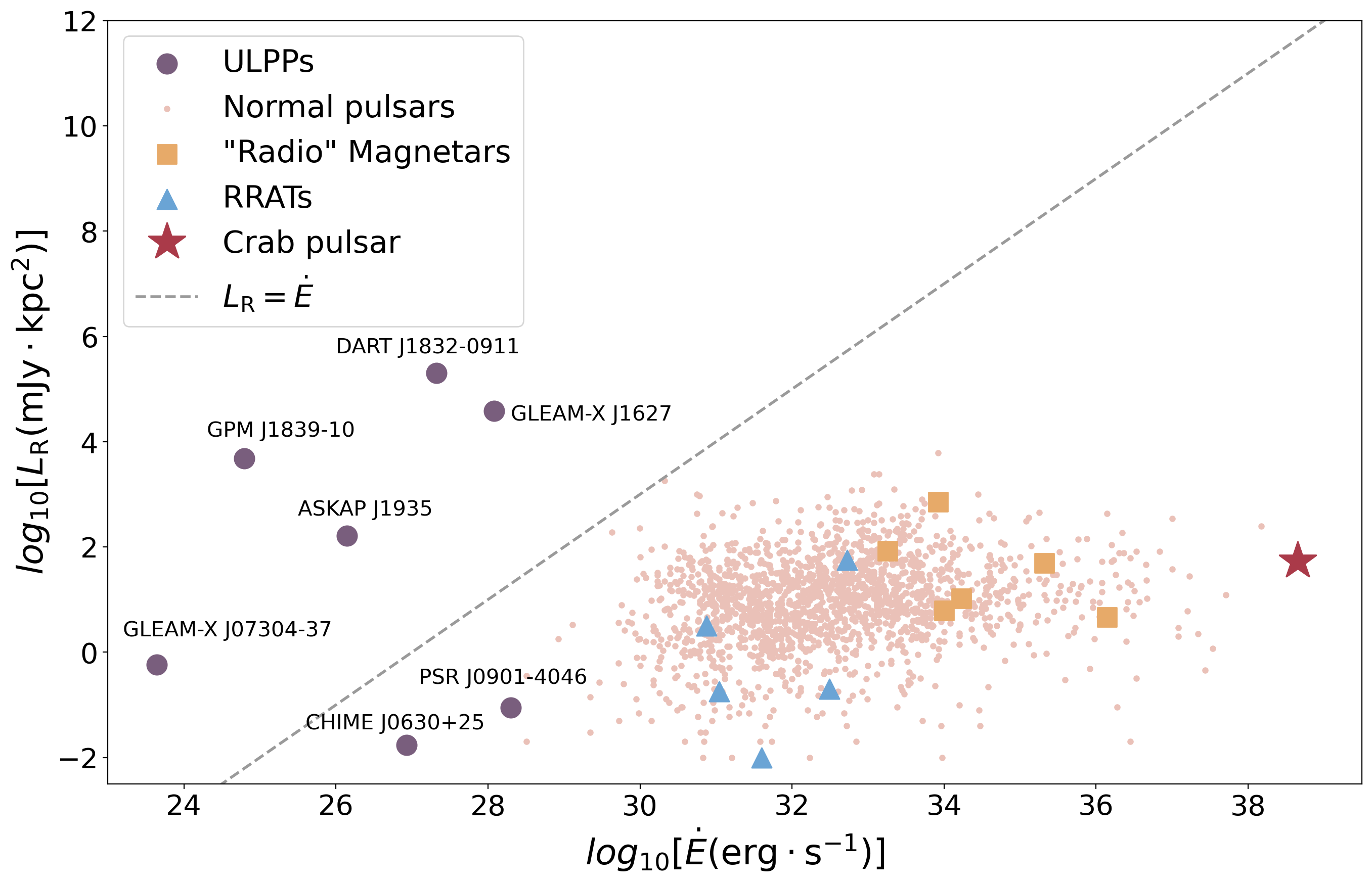}
    \caption{The distribution diagrams of $L_{R}$ and $\dot{E}$ for ULPPs ,Normal pulsars,``Radio" Magnetar and RRATs at 1400 MHz.The five-pointed stars represents the Crab Pulsar, dots represents the Normal Pulsars, squares represents the "Radio" Magnetars(Swift J1818.0-1607, SGR 1745-2900, PSR J1622-4950, XTE J1810-197, 1E 1547.0-5408 and SGR 1935+2154)\citep{Camilo2006, Camilo2007, Rea10, Levin2012, Esposito2020,Wang2024}, triangles represents the RRATs, The circles represent GLEAM-X J0704-37, ASKAP J1935, DART J1832-0911, GPM J1839-10, GLEAM-X J1627, CHIME J0630+25 and PSR J0901-4046 respectively. Notes: GCRT J1745-3009 is not marked due to lack of measured $\dot{E}$.\\
    The data are taken from the ATNF Pulsar Catalogue\citep{Manchester05},  and the ``Radio" Magnetars refer to magnetars with radio emission. }
    \label{fig5}
\end{figure}
It can be clearly noticed from Figure \ref{fig5} that the ULPPs and normal pulsars are distributed in two distinctive  clusters, with $L_{\rm R}>\dot{E}$ and $L_{\rm R}<\dot{E}$, respectively. The observed  number  of ULPPs is relatively small (7/8 sources have values of $L_{\rm R}$ and $\dot{E}$) and their distribution in the diagram of $L_{\rm R}$ vs. $\dot{E}$ is rather scattered. In contrast, the normal pulsars have a big  data set and are distributed in a rather broad area. In addition, the ULPPs have the lower $\dot{E}$ values than those of the normal pulsars. This dichotomy of distribution in $L_{R}-\dot{E}$ implies that the two types of sources would have the different radio emission mechanisms; while the coherent  radio emissions of ULPPs can be attributed to the magnetic-reconnection triggered phenomena, and those of the normal pulsars are accounted for  the rotation-powered electric voltage in the polar cap of pulsar. In other words, the ULPPs are likely to have a different magnetic field structure from that  of the normal pulsars, which consequently leads to their different physical characteristics. It is worth noting that in Figure \ref{fig5} the ULPPs with periods of $76$\,s and $421$\,s are located within the cluster of the normal pulsars but close to the cluster boundary  of the ULPPs,  which might suggest that they are the transitional sources.  Meanwhile, we cannot rule out the possibility that the spin period  ``gap" presented in Figure \ref{fig1} is caused by the selection effect of observations. It is believed that, with the improvements of observational techniques and instrumentation, more and more ULPPs will be discovered, which would thus fill this ``gap".

\subsection{Discussion of model parameters}
\begin{figure}
    \centering
    \includegraphics[width=0.8\linewidth]{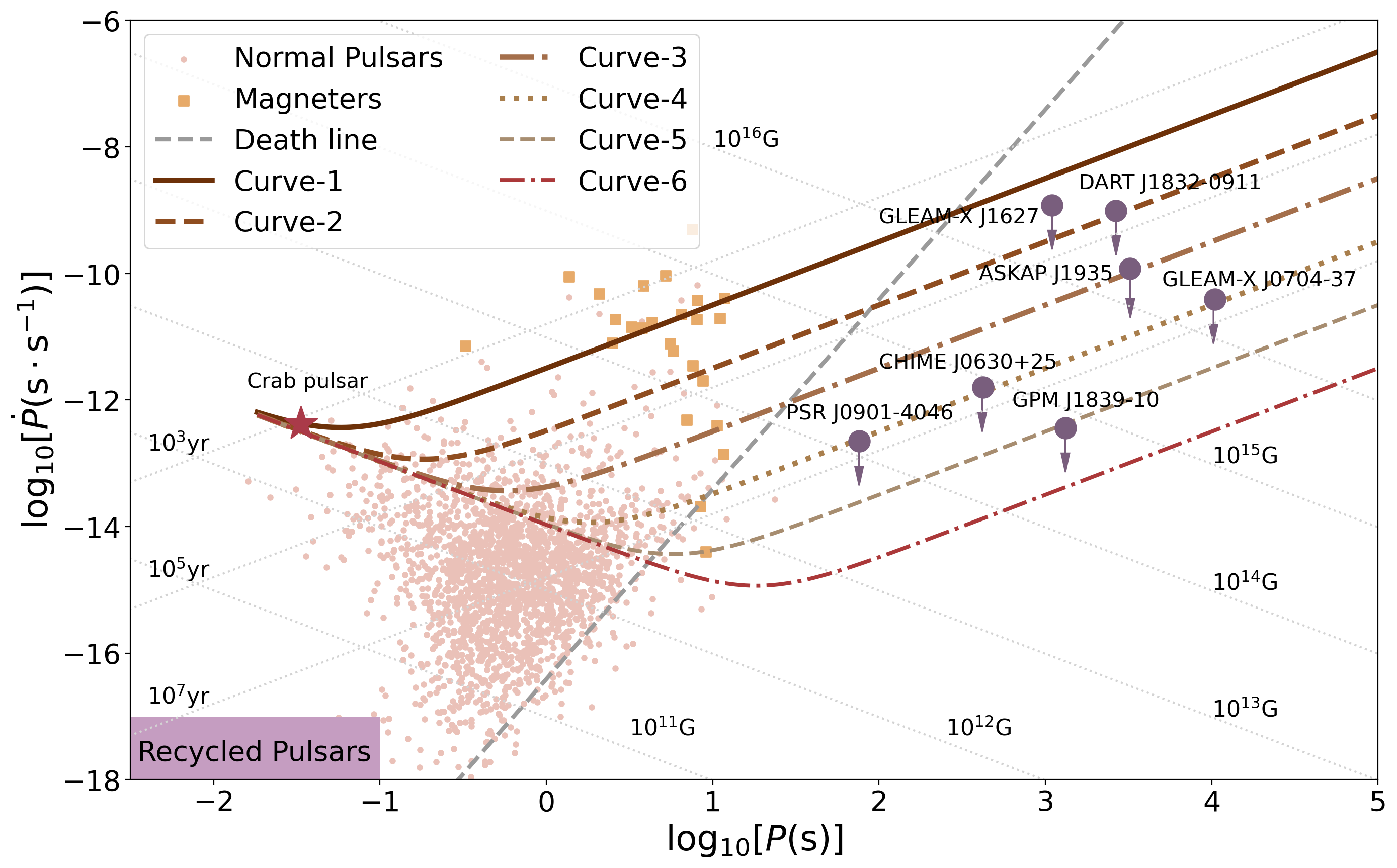}
    \caption{The period and period derivative diagram ($P-\dot{P}$ diagram) for different $b$ values. The six curves stand for the different evolution parameters of $b$ in MDRW model for the spin period evolution, with the Curve-1 ($a_{1}=2.67\times10^{-16}\,\rm {c\cdot{g}\cdot{s}}$ and $b_{1}=3.15\times10^{-12}\,\rm {c\cdot{g}\cdot{s}}$), the Curve-2 ($a_{2}=a_{1}$ and $b_{2}=0.1b_{1}$), the Curve-3 ($a_{3}=a_{1}$ and $b_{3}=0.01b_{1}$), the Curve-4 ($a_{4}=a_{1}$ and $b_{4}=0.001b_{1}$), the Curve-5 ($a_{5}=a_{1}$ and $b_{5}=0.0001b_{1}$) and the Curve-6 ($a_{6}=a_{1}$ and $b_{6}=0.00001b_{1}$). The starred sign denotes the Crab pulsar, dots represent the Normal Pulsars, squares represent the magnetars, and the circles represent GLEAM-X J0704-37, ASKAP J1935, GPM J1839-10, DART J1832-0911, GLEAM-X J1627, CHIME J0630+25 and PSR J0901-4046, respectively. Notes: GCRT J1745-3009 is not marked due to lack of measured $\dot{P}$.}
    \label{fig6}
\end{figure}
Figure \ref{fig6} shows the $P-\dot{P}$ diagram of pulsars, where curve-1 to curve-6 represent the evolutionary tracks of pulsar spin periods based on the MDRW model with the decreasing fraction of the particle flow component, i.e., the decreasing parameter $b$ values in Equation \ref{eq15}, from $b_{1}$ for curve-1 to $b_{6}=10^{-5}b_{1}$ for curve-6.
In particular, curve-1 represents the evolution of pulsar spin-down dominated by the particle flow braking, while curve-6 represents the pulsar braking evolution dominated by the magnetic dipole radiation.
It is noticed that in all of these evolutionary curves, the particle flow influence plays the major role in the later stage of the pulsar spin-down evolution. 
Take curve-6 as an example, the magnetic dipole radiation dominates the pulsar spin-down evolution before the spin period to be less than $ P\sim20\, \mathrm{s}$, and then the particle flow braking plays the major braking role after the spin period being longer than $ P\sim20\, \mathrm{s}$.
It is noticed that the MDRW model can well explain the formation and evolution of the observed spin periods of ULPPs with the appropriate parameter $b$ values, or the ratios between the magnetic dipole radiation and particle flow components.
For example, GPM J1839-10 locates the position of $P-\dot{P}$ diagram on curve-5 with the parameter value of $b_{5} = 0.0001b_{1}$, i.e., the low fraction of particle flow component.
Another example is that J0901-4046, CHIME J0630+25 and GLEAM-X J0704-37 all locate on curve-4, we suggest that these three sources share the similar fraction of particle flow component, but the different evolutionary timescales and stages.



\subsection{Model dependency}
At present, the observational sample of  ULPPs is indeed quite limited (only approximately 8 confirmed radio ULPPs have been identified). This situation gives rise to a considerable degree of uncertainty when one attempts to validate progenitor star evolution models directly on the basis of ULPP samples through statistical analysis. Despite the limited sample size, the Crab pulsar was selected as the reference source for the evolutionary starting point primarily because it is a typical representative of young pulsars. Moreover, as a standard source, the parameters of Crab pulsar have been accurately measured through long-term observations, and it has completed observational data across all wavelengths, rendering it the optimal target for testing our model. Therefore, we assume that ULPPs have the initial characteristics similar to that of Crab pulsars, so as to verify whether the model can explain the evolution of ULPPs. 
In the future, with the discovery of more ULPPs, it will deepen the understanding of the formation mechanisms of ULPPs.

\section{Conclusions}\label{sec:5}
In summary, we investigate the possible origins and emission mechanisms for the ULPPs, which are ascribed to the reconnection of the local superstrong magnetic field lines, and a combination of the dipolar slow-down of a normal pulsar and an effective particle wind braking for the long-term spin evolution. 
Unlike the normal pulsars that are the rotation-powered sources, all ULPPs with known period derivative are far below the death-line in the diagram of period versus period derivative, where the radio luminosity of 5 ULPPs are higher than their rotational energy loss rates, implying the coherent radio emissions triggered by the reconnection of the local magnetar-like field lines. 
Thus, we are inclined to think that there exist  two types of the coherent radio emission mechanisms, the magnetic reconnection induced \citep{Melrose1980, Melrose1980book, Bailes2022} and rotation-powered \citep{Gold68, Ruderman75}.  
If the ULPPs are evolved from the Crab-like pulsars with the local magnetar-like field structure, the relativistic particle wind flow can bring the NS spin period of several tens of milliseconds into the $54$\,min ULPP (ASKAP J1935) in a time scale of about  (0.1 - 1) Myr. 
The two components of magnetic fields of the ULPPs are suggested, the large scale dipolar field and local multiple field, which are corresponding to, respective,  
the coherent radio emissions of the normal pulsars by the rotation-powered electric voltage and of the ULPPs by the magnetic reconnection. 
Finally, we briefly summarize the main conclusions of this paper in the following.

1. In order for a normal pulsar with the local magnetar-like field to evolve into a ULPP, it is imperative that a particle wind  flow is involved, since the pure MDR model with the low efficiency spin-down braking  cannot account for the periods of ULPPs.

2. We have developed a model of the local superstrong magnetic field of NS 
to account for the coherent radio emissions of ULPPs, based on the magnetic-reconnection, which could provide a new route to understand the mechanism of radio emission generation for the ULPPs.

3. In general, the mechanism by which the ULPPs produce the coherent radio emission  should differ from the normal pulsars. Furthermore,  the RRATs, ``radio'' magnetars and normal pulsars, together with 2 ``fast'' ULPPs (with periods of $76$\,s and $421$\,s)  satisfy the condition of the  radio luminosity being less than the rotational energy loss rate, which may be involved in both the rotation-powered and magnetic-reconnection triggered  the coherent radio emission mechanisms.

\bibliography{sn-bibliography}

\begin{thebibliography}{94}
\providecommand{\natexlab}[1]{#1}
\providecommand{\url}[1]{{#1}}
\providecommand{\urlprefix}{URL }
\providecommand{\doi}[1]{\url{https://doi.org/#1}}
\providecommand{\eprint}[2][]{\url{#2}}
 \bibcommenthead

\bibitem[{{Afonina} et~al.(2024){Afonina}, {Biryukov}, and {Popov}}]{Afonina2024}
{Afonina} MD, {Biryukov} AV, {Popov} SB (2024) {Early accretion onset in long-period isolated pulsars}. \pasa 41:e014. \doi{10.1017/pasa.2024.12}, {\href{https://arxiv.org/abs/2310.14844}{{arXiv:2310.14844}}} {[astro-ph.HE]}

\bibitem[{{Aguilera} et~al.(2008){Aguilera}, {Pons}, and {Miralles}}]{Aguilera2008}
{Aguilera} DN, {Pons} JA, {Miralles} JA (2008) {The Impact of Magnetic Field on the Thermal Evolution of Neutron Stars}. \apjl 673(2):L167. \doi{10.1086/527547}, {\href{https://arxiv.org/abs/0712.1353}{{arXiv:0712.1353}}} {[astro-ph]}

\bibitem[{{Araujo} et~al.(2024){Araujo}, {De Lorenci}, {Peter}, and {Ruiz}}]{Araujo2024}
{Araujo} ECA, {De Lorenci} VA, {Peter} P, et~al (2024) {A phenomenological wobbling model for isolated pulsars and the braking index}. \mnras 527(3):7956--7964. \doi{10.1093/mnras/stad3531}

\bibitem[{{Arons}(2000)}]{Arons00}
{Arons} J (2000) {Pulsar Death at an Advanced Age}. In: {Kramer} M, {Wex} N, {Wielebinski} R (eds) IAU Colloq. 177: Pulsar Astronomy - 2000 and Beyond, p 449, \doi{10.48550/arXiv.astro-ph/9911478}, {\href{https://arxiv.org/abs/astro-ph/9911478}{{arXiv:astro-ph/9911478}}}

\bibitem[{{Asseo} and {Khechinashvili}(2002)}]{Asseo2002}
{Asseo} E, {Khechinashvili} D (2002) {The role of multipolar magnetic fields in pulsar magnetospheres}. \mnras 334(4):743--759. \doi{10.1046/j.1365-8711.2002.05481.x}, {\href{https://arxiv.org/abs/astro-ph/0203129}{{arXiv:astro-ph/0203129}}} {[astro-ph]}

\bibitem[{{Bailes}(2022)}]{Bailes2022}
{Bailes} M (2022) {The discovery and scientific potential of fast radio bursts}. Science 378(6620):abj3043. \doi{10.1126/science.abj3043}, {\href{https://arxiv.org/abs/2211.06048}{{arXiv:2211.06048}}} {[astro-ph.HE]}

\bibitem[{Becker(2009)}]{Becker09}
Becker W (2009) {Neutron Stars and Pulsars}, Astrophysics and Space Science Library, vol 357. Springer Berlin Heidelberg, \doi{10.1007/978-3-540-76965-1}

\bibitem[{{Beniamini} et~al.(2023){Beniamini}, {Wadiasingh}, {Hare}, {Rajwade}, {Younes}, and {van der Horst}}]{Beniamini2023}
{Beniamini} P, {Wadiasingh} Z, {Hare} J, et~al (2023) {Evidence for an abundant old population of Galactic ultra-long period magnetars and implications for fast radio bursts}. \mnras 520(2):1872--1894. \doi{10.1093/mnras/stad208}, {\href{https://arxiv.org/abs/2210.09323}{{arXiv:2210.09323}}} {[astro-ph.HE]}

\bibitem[{{Beskin} et~al.(2021){Beskin}, {Zagorulia}, and {Istomin}}]{beskin21}
{Beskin} VS, {Zagorulia} DS, {Istomin} AY (2021) {Magnetic Fields of Neutron Stars}. Astronomy Letters 47(10):686--694. \doi{10.1134/S1063773721100017}

\bibitem[{{Bhattacharya} and {Datta}(1996)}]{Bhattacharya1996}
{Bhattacharya} D, {Datta} B (1996) {Ohmic decay of magnetic flux expelled from neutron star interiors}. \mnras 282(3):1059--1063. \doi{10.1093/mnras/282.3.1059}

\bibitem[{{Bhattacharya} and {van den Heuvel}(1991)}]{Bhattacharya91}
{Bhattacharya} D, {van den Heuvel} EPJ (1991) {Formation and evolution of binary and millisecond radio pulsars}. \physrep 203(1-2):1--124. \doi{10.1016/0370-1573(91)90064-S}

\bibitem[{{Caleb} et~al.(2022){Caleb}, {Heywood}, {Rajwade}, {Malenta}, {Stappers}, {Barr}, {Chen}, {Morello}, {Sanidas}, {van den Eijnden}, {Kramer}, {Buckley}, {Brink}, {Motta}, {Woudt}, {Weltevrede}, {Jankowski}, {Surnis}, {Buchner}, {Bezuidenhout}, {Driessen}, and {Fender}}]{Caleb22}
{Caleb} M, {Heywood} I, {Rajwade} K, et~al (2022) {Discovery of a radio-emitting neutron star with an ultra-long spin period of 76 s}. Nature Astronomy 6:828--836. \doi{10.1038/s41550-022-01688-x}, {\href{https://arxiv.org/abs/2206.01346}{{arXiv:2206.01346}}} {[astro-ph.HE]}

\bibitem[{{Caleb} et~al.(2024){Caleb}, {Lenc}, {Kaplan}, {Murphy}, {Men}, {Shannon}, {Ferrario}, {Rajwade}, {Clarke}, {Giacintucci}, {Hurley-Walker}, {Hyman}, {Lower}, {McSweeney}, {Ravi}, {Barr}, {Buchner}, {Flynn}, {Hessels}, {Kramer}, {Pritchard}, and {Stappers}}]{Caleb2024}
{Caleb} M, {Lenc} E, {Kaplan} DL, et~al (2024) {An emission-state-switching radio transient with a 54-minute period}. Nature Astronomy 8:1159--1168. \doi{10.1038/s41550-024-02277-w}, {\href{https://arxiv.org/abs/2407.12266}{{arXiv:2407.12266}}} {[astro-ph.HE]}

\bibitem[{{Camilo} et~al.(2006){Camilo}, {Ransom}, {Halpern}, {Reynolds}, {Helfand}, {Zimmerman}, and {Sarkissian}}]{Camilo2006}
{Camilo} F, {Ransom} SM, {Halpern} JP, et~al (2006) {Transient pulsed radio emission from a magnetar}. \nat 442(7105):892--895. \doi{10.1038/nature04986}, {\href{https://arxiv.org/abs/astro-ph/0605429}{{arXiv:astro-ph/0605429}}} {[astro-ph]}

\bibitem[{{Camilo} et~al.(2007){Camilo}, {Ransom}, {Halpern}, and {Reynolds}}]{Camilo2007}
{Camilo} F, {Ransom} SM, {Halpern} JP, et~al (2007) {1E 1547.0-5408: A Radio-emitting Magnetar with a Rotation Period of 2 Seconds}. \apjl 666(2):L93--L96. \doi{10.1086/521826}, {\href{https://arxiv.org/abs/0708.0002}{{arXiv:0708.0002}}} {[astro-ph]}

\bibitem[{{Chen} and {Ruderman}(1993)}]{Chen1993ApJ}
{Chen} K, {Ruderman} M (1993) {Pulsar Death Lines and Death Valley}. \apj 402:264. \doi{10.1086/172129}

\bibitem[{{Cooper} and {Wadiasingh}(2024)}]{Cooper2024}
{Cooper} AJ, {Wadiasingh} Z (2024) {Beyond the Rotational Deathline: Radio Emission from Ultra-long Period Magnetars}. \mnras 533(2):2133--2155. \doi{10.1093/mnras/stae1813}, {\href{https://arxiv.org/abs/2406.04135}{{arXiv:2406.04135}}} {[astro-ph.HE]}

\bibitem[{{Cordes} and {Chatterjee}(2019)}]{Cordes19}
{Cordes} JM, {Chatterjee} S (2019) {Fast Radio Bursts: An Extragalactic Enigma}. \araa 57:417--465. \doi{10.1146/annurev-astro-091918-104501} {[astro-ph.HE]}

\bibitem[{{de Gouveia Dal Pino} and {Lazarian}(2000)}]{De2000}
{de Gouveia Dal Pino} EM, {Lazarian} A (2000) {Ultra-High-Energy Cosmic-Ray Acceleration by Magnetic Reconnection in Newborn Accretion-induced Collapse Pulsars}. \apjl 536(1):L31--L34. \doi{10.1086/312730}, {\href{https://arxiv.org/abs/astro-ph/0002155}{{arXiv:astro-ph/0002155}}} {[astro-ph]}

\bibitem[{{De Sarkar} et~al.(2022){De Sarkar}, {Zhang}, {Mart{\'\i}n}, {Torres}, {Li}, and {Hou}}]{De2022}
{De Sarkar} A, {Zhang} W, {Mart{\'\i}n} J, et~al (2022) {LHAASO J2226+6057 as a pulsar wind nebula}. \aap 668:A23. \doi{10.1051/0004-6361/202244841}, {\href{https://arxiv.org/abs/2209.13285}{{arXiv:2209.13285}}} {[astro-ph.HE]}

\bibitem[{{Dong} et~al.(2024){Dong}, {Clarke}, {Curtin}, {Kumar}, {Stairs}, {Chatterjee}, {Cook}, {Fonseca}, {Gaensler}, {Hessels}, {Kaspi}, {Lazda}, {Masui}, {McKee}, {Meyers}, {Pearlman}, {Ransom}, {Scholz}, {Shin}, {Smith}, and {Tan}}]{Dong2024}
{Dong} FA, {Clarke} T, {Curtin} AP, et~al (2024) {The discovery of a nearby 421\raisebox{-0.5ex}\textasciitilde transient with CHIME/FRB/Pulsar}. arXiv e-prints arXiv:2407.07480. \doi{10.48550/arXiv.2407.07480}, {\href{https://arxiv.org/abs/2407.07480}{{arXiv:2407.07480}}} {[astro-ph.HE]}

\bibitem[{{Duncan} and {Thompson}(1992)}]{Ducan1992}
{Duncan} RC, {Thompson} C (1992) {Formation of Very Strongly Magnetized Neutron Stars: Implications for Gamma-Ray Bursts}. \apjl 392:L9. \doi{10.1086/186413}

\bibitem[{{Ek{\c{s}}i} et~al.(2016){Ek{\c{s}}i}, {Anda{\c{c}}}, {{\c{C}}{\i}k{\i}nto{\u{g}}lu}, {G{\"u}gercino{\u{g}}lu}, {Vahdat Motlagh}, and {K{\i}z{\i}ltan}}]{Eksi16}
{Ek{\c{s}}i} KY, {Anda{\c{c}}} IC, {{\c{C}}{\i}k{\i}nto{\u{g}}lu} S, et~al (2016) {The Inclination Angle and Evolution of the Braking Index of Pulsars with Plasma-filled Magnetosphere: Application to the High Braking Index of PSR J1640-4631}. \apj 823(1):34. \doi{10.3847/0004-637X/823/1/34}, {\href{https://arxiv.org/abs/1603.01487}{{arXiv:1603.01487}}} {[astro-ph.HE]}

\bibitem[{{Esposito} et~al.(2020){Esposito}, {Rea}, {Borghese}, {Coti Zelati}, {Vigan{\`o}}, {Israel}, {Tiengo}, {Ridolfi}, {Possenti}, {Burgay}, {G{\"o}tz}, {Pintore}, {Stella}, {Dehman}, {Ronchi}, {Campana}, {Garcia-Garcia}, {Graber}, {Mereghetti}, {Perna}, {Rodr{\'\i}guez Castillo}, {Turolla}, and {Zane}}]{Esposito2020}
{Esposito} P, {Rea} N, {Borghese} A, et~al (2020) {A Very Young Radio-loud Magnetar}. \apjl 896(2):L30. \doi{10.3847/2041-8213/ab9742}, {\href{https://arxiv.org/abs/2004.04083}{{arXiv:2004.04083}}} {[astro-ph.HE]}

\bibitem[{{Esposito} et~al.(2021){Esposito}, {Rea}, and {Israel}}]{Esposito2021}
{Esposito} P, {Rea} N, {Israel} GL (2021) {Magnetars: A Short Review and Some Sparse Considerations}. In: {Belloni} TM, {M{\'e}ndez} M, {Zhang} C (eds) Timing Neutron Stars: Pulsations, Oscillations and Explosions, pp 97--142, \doi{10.1007/978-3-662-62110-3_3}, {\href{https://arxiv.org/abs/1803.05716}{{arXiv:1803.05716}}}

\bibitem[{{Fender} and {Hendry}(2000)}]{Fender2000}
{Fender} RP, {Hendry} MA (2000) {The radio luminosity of persistent X-ray binaries}. \mnras 317(1):1--8. \doi{10.1046/j.1365-8711.2000.03443.x}, {\href{https://arxiv.org/abs/astro-ph/0001502}{{arXiv:astro-ph/0001502}}} {[astro-ph]}

\bibitem[{{Gen{\c{c}}ali} et~al.(2022){Gen{\c{c}}ali}, {Ertan}, and {Alpar}}]{Gencali2022}
{Gen{\c{c}}ali} AA, {Ertan} {\"U}, {Alpar} MA (2022) {Evolution of the long-period pulsar GLEAM-X J162759.5-523504.3}. \mnras 513(1):L68--L71. \doi{10.1093/mnrasl/slac034}, {\href{https://arxiv.org/abs/2202.06852}{{arXiv:2202.06852}}} {[astro-ph.HE]}

\bibitem[{{Geppert} and {Vigan{\`o}}(2014)}]{Geppert2014}
{Geppert} U, {Vigan{\`o}} D (2014) {Creation of magnetic spots at the neutron star surface}. \mnras 444(4):3198--3208. \doi{10.1093/mnras/stu1675}, {\href{https://arxiv.org/abs/1408.3833}{{arXiv:1408.3833}}} {[astro-ph.SR]}

\bibitem[{{Geppert} et~al.(2013){Geppert}, {Gil}, and {Melikidze}}]{Geppert2013}
{Geppert} U, {Gil} J, {Melikidze} G (2013) {Radio pulsar activity and the crustal Hall drift}. \mnras 435(4):3262--3271. \doi{10.1093/mnras/stt1527}, {\href{https://arxiv.org/abs/1308.2718}{{arXiv:1308.2718}}} {[astro-ph.SR]}

\bibitem[{{Giraud} and {P{\'e}tri}(2020)}]{Giraud2020}
{Giraud} Q, {P{\'e}tri} J (2020) {Radio and high-energy emission of pulsars revealed by general relativity}. \aap 639:A75. \doi{10.1051/0004-6361/202037979}

\bibitem[{{Gold}(1968)}]{Gold68}
{Gold} T (1968) {Rotating Neutron Stars as the Origin of the Pulsating Radio Sources}. \nat 218(5143):731--732. \doi{10.1038/218731a0}

\bibitem[{{Goldreich} and {Reisenegger}(1992)}]{Goldreich1992}
{Goldreich} P, {Reisenegger} A (1992) {Magnetic Field Decay in Isolated Neutron Stars}. \apj 395:250. \doi{10.1086/171646}

\bibitem[{{Gourgouliatos} and {Cumming}(2015)}]{Gourgouliatos2015}
{Gourgouliatos} KN, {Cumming} A (2015) {Hall drift and the braking indices of young pulsars}. \mnras 446(1):1121--1128. \doi{10.1093/mnras/stu2140}, {\href{https://arxiv.org/abs/1406.3640}{{arXiv:1406.3640}}} {[astro-ph.SR]}

\bibitem[{{G{\"u}del} et~al.(2008){G{\"u}del}, {Briggs}, {Montmerle}, {Audard}, {Rebull}, and {Skinner}}]{Gudel2008}
{G{\"u}del} M, {Briggs} KR, {Montmerle} T, et~al (2008) {Million-Degree Plasma Pervading the Extended Orion Nebula}. Science 319(5861):309. \doi{10.1126/science.1149926}, {\href{https://arxiv.org/abs/0712.0476}{{arXiv:0712.0476}}} {[astro-ph]}

\bibitem[{{Haensel} et~al.(1990){Haensel}, {Denisov}, and {Popov}}]{Haensel1990}
{Haensel} P, {Denisov} A, {Popov} S (1990) {Neutron star corequake implied by pion condensation - Dynamic, neutrino and thermal effects}. \aap 240(1):78--84

\bibitem[{Haensel et~al.(2007)Haensel, Potekhin, and Yakovlev}]{Haensel07}
Haensel P, Potekhin AY, Yakovlev DG (2007) Neutron Stars 1. Springer

\bibitem[{{Hurley-Walker} et~al.(2022){Hurley-Walker}, {Zhang}, {Bahramian}, {McSweeney}, {O'Doherty}, {Hancock}, {Morgan}, {Anderson}, {Heald}, and {Galvin}}]{Hurley22}
{Hurley-Walker} N, {Zhang} X, {Bahramian} A, et~al (2022) {A radio transient with unusually slow periodic emission}. \nat 601(7894):526--530. \doi{10.1038/s41586-021-04272-x}

\bibitem[{{Hurley-Walker} et~al.(2023){Hurley-Walker}, {Rea}, {McSweeney}, {Meyers}, {Lenc}, {Heywood}, {Hyman}, {Men}, {Clarke}, {Coti Zelati}, {Price}, {Horv{\'a}th}, {Galvin}, {Anderson}, {Bahramian}, {Barr}, {Bhat}, {Caleb}, {Dall'Ora}, {de Martino}, {Giacintucci}, {Morgan}, {Rajwade}, {Stappers}, and {Williams}}]{Hurley23}
{Hurley-Walker} N, {Rea} N, {McSweeney} SJ, et~al (2023) {A long-period radio transient active for three decades}. \nat 619(7970):487--490. \doi{10.1038/s41586-023-06202-5}

\bibitem[{{Hurley-Walker} et~al.(2024){Hurley-Walker}, {McSweeney}, {Bahramian}, {Rea}, {Horv{\'a}th}, {Buchner}, {Williams}, {Meyers}, {Strader}, {Aydi}, {Urquhart}, {Chomiuk}, {Galvin}, {Coti Zelati}, and {Bailes}}]{Hurley-Walker24}
{Hurley-Walker} N, {McSweeney} SJ, {Bahramian} A, et~al (2024) {A 2.9 hr Periodic Radio Transient with an Optical Counterpart}. \apjl 976(2):L21. \doi{10.3847/2041-8213/ad890e}, {\href{https://arxiv.org/abs/2408.15757}{{arXiv:2408.15757}}} {[astro-ph.SR]}

\bibitem[{{Hyman} et~al.(2005){Hyman}, {Lazio}, {Kassim}, {Ray}, {Markwardt}, and {Yusef-Zadeh}}]{Hyman05}
{Hyman} SD, {Lazio} TJW, {Kassim} NE, et~al (2005) {A powerful bursting radio source towards the Galactic Centre}. \nat 434(7029):50--52. \doi{10.1038/nature03400}, {\href{https://arxiv.org/abs/astro-ph/0503052}{{arXiv:astro-ph/0503052}}} {[astro-ph]}

\bibitem[{{Igoshev} et~al.(2021){Igoshev}, {Popov}, and {Hollerbach}}]{Igoshev2021}
{Igoshev} AP, {Popov} SB, {Hollerbach} R (2021) {Evolution of Neutron Star Magnetic Fields}. Universe 7(9):351. \doi{10.3390/universe7090351}, {\href{https://arxiv.org/abs/2109.05584}{{arXiv:2109.05584}}} {[astro-ph.HE]}

\bibitem[{{Kaspi} and {Beloborodov}(2017)}]{Kaspi17}
{Kaspi} VM, {Beloborodov} AM (2017) {Magnetars}. \araa 55(1):261--301. \doi{10.1146/annurev-astro-081915-023329} {[astro-ph.HE]}

\bibitem[{{Katz}(2022)}]{Katz2022}
{Katz} JI (2022) {GLEAM-X J162759.5‑523504.3 as a white dwarf pulsar}. \apss 367(11):108. \doi{10.1007/s10509-022-04146-2}, {\href{https://arxiv.org/abs/2203.08112}{{arXiv:2203.08112}}} {[astro-ph.SR]}

\bibitem[{{Kontorovich}(2010)}]{Kontorovich2010}
{Kontorovich} VM (2010) {Electromagnetic tornado in the vacuum gap of a pulsar}. Soviet Journal of Experimental and Theoretical Physics 110(6):966--972. \doi{10.1134/S1063776110060051}

\bibitem[{{Lai} and {Qian}(1998)}]{LaiApJ}
{Lai} D, {Qian} YZ (1998) {Neutrino Transport in Strongly Magnetized Proto-Neutron Stars and the Origin of Pulsar Kicks: The Effect of Asymmetric Magnetic Field Topology}. \apj 505(2):844--853. \doi{10.1086/306203}, {\href{https://arxiv.org/abs/astro-ph/9802345}{{arXiv:astro-ph/9802345}}} {[astro-ph]}

\bibitem[{{Levin} et~al.(2012){Levin}, {Bailes}, {Bates}, {Bhat}, {Burgay}, {Burke-Spolaor}, {D'Amico}, {Johnston}, {Keith}, {Kramer}, {Milia}, {Possenti}, {Stappers}, and {van Straten}}]{Levin2012}
{Levin} L, {Bailes} M, {Bates} SD, et~al (2012) {Radio emission evolution, polarimetry and multifrequency single pulse analysis of the radio magnetar PSR J1622-4950}. \mnras 422(3):2489--2500. \doi{10.1111/j.1365-2966.2012.20807.x}, {\href{https://arxiv.org/abs/1204.2045}{{arXiv:1204.2045}}} {[astro-ph.HE]}

\bibitem[{Li et~al.(2024)Li, Yuan, Wu, Yan, Lv, Tsai, Wang, Zhu, Deng, Lan, Xu, Chen, Meng, Li, Li, Zhou, Yang, Xue, Lu, Miao, Wang, Niu, Fang, Fu, Feng, Zhang, Jiang, Miao, Chen, Sun, Yang, Deng, Dai, Chen, Yao, Liu, Li, Zhang, Yang, Zhou, Yiyizhou, Zhang, Niu, Zhao, Zhang, Peng, Wu, and Wang}]{Li2024}
Li D, Yuan M, Wu L, et~al (2024) A 44-minute periodic radio transient in a supernova remnant. \urlprefix\url{https://arxiv.org/abs/2411.15739}, {\href{https://arxiv.org/abs/2411.15739}{{arXiv:2411.15739}}}

\bibitem[{{Lorimer} and {Kramer}(2012)}]{Lorimer12}
{Lorimer} DR, {Kramer} M (2012) {Handbook of Pulsar Astronomy}. Cambridge University Press

\bibitem[{{Lu} and {Kumar}(2018)}]{Lu18}
{Lu} W, {Kumar} P (2018) {On the radiation mechanism of repeating fast radio bursts}. \mnras 477(2):2470--2493. \doi{10.1093/mnras/sty716}, {\href{https://arxiv.org/abs/1710.10270}{{arXiv:1710.10270}}} {[astro-ph.HE]}

\bibitem[{{Lyne} and {Graham-Smith}(2022)}]{lyne2022}
{Lyne} A, {Graham-Smith} FSB (2022) {Pulsar Astronomy}

\bibitem[{{Lyne} et~al.(2015){Lyne}, {Jordan}, {Graham-Smith}, {Espinoza}, {Stappers}, and {Weltevrede}}]{Lyne15}
{Lyne} AG, {Jordan} CA, {Graham-Smith} F, et~al (2015) {45 years of rotation of the Crab pulsar}. \mnras 446(1):857--864. \doi{10.1093/mnras/stu2118} {[astro-ph.HE]}

\bibitem[{{Manchester} and {Taylor}(1977)}]{Manchester77}
{Manchester} RN, {Taylor} JH (1977) {Pulsars}. W. H. Freeman

\bibitem[{{Manchester} et~al.(2005){Manchester}, {Hobbs}, {Teoh}, and {Hobbs}}]{Manchester05}
{Manchester} RN, {Hobbs} GB, {Teoh} A, et~al (2005) {The Australia Telescope National Facility Pulsar Catalogue}. \aj 129(4):1993--2006. \doi{10.1086/428488} {[astro-ph]}

\bibitem[{{Melrose}(1980{\natexlab{a}})}]{Melrose1980}
{Melrose} DB (1980{\natexlab{a}}) {A Plasma Emission Mechanism for Type-I Solar Radio Emission}. Solar Physics 67(2):357--375. \doi{10.1007/BF00149813}

\bibitem[{{Melrose}(1980{\natexlab{b}})}]{Melrose1980book}
{Melrose} DB (1980{\natexlab{b}}) {Plasma astrohysics. Nonthermal processes in diffuse magnetized plasmas - Vol.1: The emission, absorption and transfer of waves in plasmas; Vol.2: Astrophysical applications}

\bibitem[{{Melrose}(1997)}]{Melrose1997}
{Melrose} DB (1997) {A Solar Flare Model Based on Magnetic Reconnection between Current-carrying Loops}. \apj 486(1):521--533. \doi{10.1086/304521}

\bibitem[{{Melrose} and {Yuen}(2016)}]{Melrose16}
{Melrose} DB, {Yuen} R (2016) {Pulsar electrodynamics: an unsolved problem}. Journal of Plasma Physics 82(2):635820202. \doi{10.1017/S0022377816000398}, {\href{https://arxiv.org/abs/1604.03623}{{arXiv:1604.03623}}} {[astro-ph.HE]}

\bibitem[{{Meszaros}(1992)}]{Meszaros92}
{Meszaros} P (1992) {High-energy radiation from magnetized neutron stars}. Chicago: University of Chicago Press

\bibitem[{{Michel}(1969)}]{Michel69a}
{Michel} FC (1969) {Relativistic Stellar-Wind Torques}. \apj 158:727. \doi{10.1086/150233}

\bibitem[{{Michel}(1982)}]{Michel82}
{Michel} FC (1982) {Theory of pulsar magnetospheres}. Reviews of Modern Physics 54(1):1--66. \doi{10.1103/RevModPhys.54.1}

\bibitem[{{Michel}(1994)}]{Michel94}
{Michel} FC (1994) {Magnetic Structure of Pulsar Winds}. \apj 431:397. \doi{10.1086/174493}

\bibitem[{{Miller} and {Miller}(2015)}]{Miller15}
{Miller} MC, {Miller} JM (2015) {The masses and spins of neutron stars and stellar-mass black holes}. \physrep 548:1--34. \doi{10.1016/j.physrep.2014.09.003}, {\href{https://arxiv.org/abs/1408.4145}{{arXiv:1408.4145}}} {[astro-ph.HE]}

\bibitem[{{Most} and {Philippov}(2023)}]{MOST2023}
{Most} ER, {Philippov} AA (2023) {Reconnection-Powered Fast Radio Transients from Coalescing Neutron Star Binaries}. Physical Review Letters 130(24):245201. \doi{10.1103/PhysRevLett.130.245201}, {\href{https://arxiv.org/abs/2207.14435}{{arXiv:2207.14435}}} {[astro-ph.HE]}

\bibitem[{{Muslimov} and {Page}(1996)}]{Muslimov1996}
{Muslimov} A, {Page} D (1996) {Magnetic and Spin History of Very Young Pulsars}. \apj 458:347. \doi{10.1086/176817}, {\href{https://arxiv.org/abs/astro-ph/9505116}{{arXiv:astro-ph/9505116}}} {[astro-ph]}

\bibitem[{{Olausen} and {Kaspi}(2014)}]{Olausen2014}
{Olausen} SA, {Kaspi} VM (2014) {The McGill Magnetar Catalog}. \apjs 212(1):6. \doi{10.1088/0067-0049/212/1/6}, {\href{https://arxiv.org/abs/1309.4167}{{arXiv:1309.4167}}} {[astro-ph.HE]}

\bibitem[{{Parker}(1957)}]{Parker1957}
{Parker} EN (1957) {Sweet's Mechanism for Merging Magnetic Fields in Conducting Fluids}. Journal of Geophysical Research 62(4):509--520. \doi{10.1029/JZ062i004p00509}

\bibitem[{{Pelisoli} et~al.(2024){Pelisoli}, {Chomiuk}, {Strader}, {Marsh}, {Aydi}, {Dage}, {Kyer}, {Molina}, {Panurach}, {Urquhart}, {Maccarone}, {Rich}, {Rodriguez}, {Breedt}, {Brown}, {Dhillon}, {Dyer}, {Gaensicke}, {Garbutt}, {Green}, {Kennedy}, {Kerry}, {Littlefair}, {Munday}, and {Parsons}}]{Pelisoli2024}
{Pelisoli} I, {Chomiuk} L, {Strader} J, et~al (2024) {A survey for radio emission from white dwarfs in the VLA Sky Survey}. \mnras 531(1):1805--1822. \doi{10.1093/mnras/stae1275}, {\href{https://arxiv.org/abs/2402.11015}{{arXiv:2402.11015}}} {[astro-ph.SR]}

\bibitem[{{P{\'e}tri}(2015)}]{Petri15}
{P{\'e}tri} J (2015) {Multipolar electromagnetic fields around neutron stars: exact vacuum solutions and related properties}. \mnras 450(1):714--742. \doi{10.1093/mnras/stv598}, {\href{https://arxiv.org/abs/1503.05307}{{arXiv:1503.05307}}} {[astro-ph.HE]}

\bibitem[{{P{\'e}tri}(2019)}]{Petri19}
{P{\'e}tri} J (2019) {The illusion of neutron star magnetic field estimates}. \mnras 485(4):4573--4587. \doi{10.1093/mnras/stz711}, {\href{https://arxiv.org/abs/1903.01528}{{arXiv:1903.01528}}} {[astro-ph.HE]}

\bibitem[{{P{\'e}tri}(2020)}]{petri2020}
{P{\'e}tri} J (2020) {Electrodynamics and Radiation from Rotating Neutron Star Magnetospheres}. Universe 6(1):15. \doi{10.3390/universe6010015}, {\href{https://arxiv.org/abs/2001.03422}{{arXiv:2001.03422}}} {[astro-ph.HE]}

\bibitem[{{Philippov} and {Kramer}(2022)}]{philippov2022}
{Philippov} A, {Kramer} M (2022) {Pulsar Magnetospheres and Their Radiation}. \araa 60:495--558. \doi{10.1146/annurev-astro-052920-112338}

\bibitem[{{Pons} and {Vigan{\`o}}(2019)}]{Pons2019}
{Pons} JA, {Vigan{\`o}} D (2019) {Magnetic, thermal and rotational evolution of isolated neutron stars}. Living Reviews in Computational Astrophysics 5(1):3. \doi{10.1007/s41115-019-0006-7}, {\href{https://arxiv.org/abs/1911.03095}{{arXiv:1911.03095}}} {[astro-ph.HE]}

\bibitem[{{Qu} and {Zhang}(2024)}]{Qu2024}
{Qu} Y, {Zhang} B (2024) {Magnetic Interaction in White Dwarf Binaries as Mechanism for Long-Period Radio Transients}. arXiv e-prints arXiv:2409.05978. \doi{10.48550/arXiv.2409.05978}, {\href{https://arxiv.org/abs/2409.05978}{{arXiv:2409.05978}}} {[astro-ph.HE]}

\bibitem[{{Rea} et~al.(2010){Rea}, {Esposito}, {Turolla}, {Israel}, {Zane}, {Stella}, {Mereghetti}, {Tiengo}, {G{\"o}tz}, {G{\"o}{\u{g}}{\"u}{\c{s}}}, and {Kouveliotou}}]{Rea10}
{Rea} N, {Esposito} P, {Turolla} R, et~al (2010) {A Low-Magnetic-Field Soft Gamma Repeater}. Science 330(6006):944. \doi{10.1126/science.1196088} {[astro-ph.HE]}

\bibitem[{{Rea} et~al.(2022){Rea}, {Coti Zelati}, {Dehman}, {Hurley-Walker}, {de Martino}, {Bahramian}, {Buckley}, {Brink}, {Kawka}, {Pons}, {Vigan{\`o}}, {Graber}, {Ronchi}, {Pardo Araujo}, {Borghese}, {Parent}, and {Galvin}}]{Rea2022}
{Rea} N, {Coti Zelati} F, {Dehman} C, et~al (2022) {Constraining the Nature of the 18 min Periodic Radio Transient GLEAM-X J162759.5-523504.3 via Multiwavelength Observations and Magneto-thermal Simulations}. \apj 940(1):72. \doi{10.3847/1538-4357/ac97ea}, {\href{https://arxiv.org/abs/2210.01903}{{arXiv:2210.01903}}} {[astro-ph.HE]}

\bibitem[{{Rezzolla} et~al.(2018){Rezzolla}, {Pizzochero}, {Jones}, {Rea}, and {Vida{\~n}a}}]{Rezzolla2018}
{Rezzolla} L, {Pizzochero} P, {Jones} DI, et~al (eds) (2018) {The Physics and Astrophysics of Neutron Stars}, Astrophysics and Space Science Library, vol 457, \doi{10.1007/978-3-319-97616-7}

\bibitem[{{Ronchi} et~al.(2022){Ronchi}, {Rea}, {Graber}, and {Hurley-Walker}}]{Ronchi2022}
{Ronchi} M, {Rea} N, {Graber} V, et~al (2022) {Long-period Pulsars as Possible Outcomes of Supernova Fallback Accretion}. \apj 934(2):184. \doi{10.3847/1538-4357/ac7cec}, {\href{https://arxiv.org/abs/2201.11704}{{arXiv:2201.11704}}} {[astro-ph.HE]}

\bibitem[{{Ruderman} and {Sutherland}(1975)}]{Ruderman75}
{Ruderman} MA, {Sutherland} PG (1975) {Theory of pulsars: polar gaps, sparks, and coherent microwave radiation.} \apj 196:51--72. \doi{10.1086/153393}

\bibitem[{{Schopper} et~al.(1998){Schopper}, {Lesch}, and {Birk}}]{Schopper1998}
{Schopper} R, {Lesch} H, {Birk} GT (1998) {Magnetic reconnection and particle acceleration in active galactic nuclei}. \aap 335:26--32. \doi{10.48550/arXiv.astro-ph/9803329}, {\href{https://arxiv.org/abs/astro-ph/9803329}{{arXiv:astro-ph/9803329}}} {[astro-ph]}

\bibitem[{{Shapiro} and {Teukolsky}(1983)}]{Shapiro83}
{Shapiro} SL, {Teukolsky} SA (1983) {Black holes, white dwarfs, and neutron stars : the physics of compact objects}. A Wiley-Interscience Publication, New York: Wiley

\bibitem[{{Spitkovsky} et~al.(2002){Spitkovsky}, {Levin}, and {Ushomirsky}}]{Spitkovsky02}
{Spitkovsky} A, {Levin} Y, {Ushomirsky} G (2002) {Propagation of Thermonuclear Flames on Rapidly Rotating Neutron Stars: Extreme Weather during Type I X-Ray Bursts}. \apj 566(2):1018--1038. \doi{10.1086/338040}, {\href{https://arxiv.org/abs/astro-ph/0108074}{{arXiv:astro-ph/0108074}}} {[astro-ph]}

\bibitem[{{Suvorov} and {Melatos}(2023)}]{Suvorov2023}
{Suvorov} AG, {Melatos} A (2023) {Evolutionary implications of a magnetar interpretation for GLEAM-X J162759.5-523504.3}. \mnras 520(1):1590--1600. \doi{10.1093/mnras/stad274}, {\href{https://arxiv.org/abs/2301.08541}{{arXiv:2301.08541}}} {[astro-ph.HE]}

\bibitem[{{Thompson} and {Beloborodov}(2005)}]{Thompson2005}
{Thompson} C, {Beloborodov} AM (2005) {High-Energy Emission from Magnetars}. \apj 634(1):565--569. \doi{10.1086/432245}, {\href{https://arxiv.org/abs/astro-ph/0408538}{{arXiv:astro-ph/0408538}}} {[astro-ph]}

\bibitem[{{Tong}(2023{\natexlab{a}})}]{Tong23}
{Tong} H (2023{\natexlab{a}}) {Discussions on the Nature of GLEAM-X J162759.5-523504.3}. \apj 943(1):3. \doi{10.3847/1538-4357/aca7fa}, {\href{https://arxiv.org/abs/2204.01957}{{arXiv:2204.01957}}} {[astro-ph.HE]}

\bibitem[{{Tong}(2023{\natexlab{b}})}]{Tong2023}
{Tong} H (2023{\natexlab{b}}) {On the Nature of Long Period Radio Pulsar GPM J1839-10: Death Line and Pulse Width}. Research in Astronomy and Astrophysics 23(12):125018. \doi{10.1088/1674-4527/ad034c}, {\href{https://arxiv.org/abs/2307.14829}{{arXiv:2307.14829}}} {[astro-ph.HE]}

\bibitem[{{Uzdensky}(2011)}]{Uzdensky2011}
{Uzdensky} DA (2011) {Magnetic Reconnection in Extreme Astrophysical Environments}. Space Science Reviews 160(1-4):45--71. \doi{10.1007/s11214-011-9744-5}, {\href{https://arxiv.org/abs/1101.2472}{{arXiv:1101.2472}}} {[astro-ph.HE]}

\bibitem[{{Uzdensky}(2016)}]{Uzdensky2016}
{Uzdensky} DA (2016) {Radiative Magnetic Reconnection in Astrophysics}. In: {Gonzalez} W, {Parker} E (eds) Magnetic Reconnection: Concepts and Applications, p 473, \doi{10.1007/978-3-319-26432-5_12}, {\href{https://arxiv.org/abs/1510.05397}{{arXiv:1510.05397}}}

\bibitem[{{Wang} et~al.(2024{\natexlab{a}}){Wang}, {Li}, {Ji}, {Hou}, {G{\"u}gercino{\u{g}}lu}, {Li}, {Torres}, {Chen}, {Niu}, {Zhu}, {Zhang}, {Liang}, {Zhang}, {Ge}, {Dai}, {Lin}, {Han}, {Feng}, {Niu}, {Zhang}, {Zhou}, {Xu}, {Zhang}, {Jiang}, {Miao}, {Yuan}, {Wang}, {Zhou}, {Fang}, {Yue}, {Wu}, {Wang}, {Wang}, {Gan}, {Li}, {Sun}, {Chi}, {Zhang}, {Cao}, {Lu}, and {Wang}}]{Wang2024}
{Wang} P, {Li} J, {Ji} L, et~al (2024{\natexlab{a}}) {X-Ray Hardening Preceding the Onset of SGR 1935+2154's Radio Pulsar Phase}. \apjs 275(2):39. \doi{10.3847/1538-4365/ad7c3f}, {\href{https://arxiv.org/abs/2308.08832}{{arXiv:2308.08832}}} {[astro-ph.HE]}

\bibitem[{{Wang} et~al.(2024{\natexlab{b}}){Wang}, {Rea}, {Bao}, {Kaplan}, {Lenc}, {Wadiasingh}, {Hare}, {Zic}, {Anumarlapudi}, {Bera}, {Beniamini}, {Cooper}, {Clarke}, {Deller}, {Dawson}, {Glowacki}, {Hurley-Walker}, {McSweeney}, {Polisensky}, {Peters}, {Younes}, {Bannister}, {Caleb}, {Dage}, {James}, {Kasliwal}, {Karambelkar}, {Lower}, {Mori}, {Ocker}, {P{\'e}rez-Torres}, {Qiu}, {Rose}, {Shannon}, {Taub}, {Wang}, {Wang}, {Zhao}, {Bhat}, {Dobie}, {Driessen}, {Murphy}, {Jaini}, {Deng}, {Jahns-Schindler}, {Lee}, {Pritchard}, {Tuthill}, and {Thyagarajan}}]{WangZiteng2024}
{Wang} Z, {Rea} N, {Bao} T, et~al (2024{\natexlab{b}}) {Detection of X-ray Emission from a Bright Long-Period Radio Transient}. arXiv e-prints arXiv:2411.16606. \doi{10.48550/arXiv.2411.16606}, {\href{https://arxiv.org/abs/2411.16606}{{arXiv:2411.16606}}} {[astro-ph.HE]}

\bibitem[{{Yang} et~al.(2024){Yang}, {Li}, {Gao}, and {Xu}}]{Yang2024}
{Yang} HR, {Li} XD, {Gao} SJ, et~al (2024) {Instability in Supernova Fallback Disks and Its Effect on the Formation of Ultralong Period Pulsars}. \apj 976(1):77. \doi{10.3847/1538-4357/ad83d4}, {\href{https://arxiv.org/abs/2410.05944}{{arXiv:2410.05944}}} {[astro-ph.HE]}

\bibitem[{{Zhang}(2003)}]{zhangbing03}
{Zhang} B (2003) {Radio Pulsar Death}. Acta Astronomica Sinica 44:215--222. \doi{10.48550/arXiv.astro-ph/0209160}, {\href{https://arxiv.org/abs/astro-ph/0209160}{{arXiv:astro-ph/0209160}}} {[astro-ph]}

\bibitem[{{Zhang}(2020)}]{Zhang2020Natur}
{Zhang} B (2020) {The physical mechanisms of fast radio bursts}. \nat 587(7832):45--53. \doi{10.1038/s41586-020-2828-1}, {\href{https://arxiv.org/abs/2011.03500}{{arXiv:2011.03500}}} {[astro-ph.HE]}

\bibitem[{{Zhang} et~al.(2022){Zhang}, {Cui}, {Li}, {Wang}, {Wang}, {Wang}, {Zhang}, {Peng}, {Zhu}, {Yang}, and {Pan}}]{Zhang22}
{Zhang} CM, {Cui} XH, {Li} D, et~al (2022) {Evolution of Spin Period and Magnetic Field of the Crab Pulsar: Decay of the Braking Index by the Particle Wind Flow Torque}. Universe 8(12):628. \doi{10.3390/universe8120628}, {\href{https://arxiv.org/abs/2212.04674}{{arXiv:2212.04674}}} {[astro-ph.HE]}

\bibitem[{{Zhang}(2023)}]{zhang2023}
{Zhang} H (2023) {Solar Magnetism}. \doi{10.1007/978-981-99-1759-4}

\end{thebibliography}

\section*{Author contributions}

Zhi-Yao Yang and Cheng-Min Zhang wrote the main manuscript text, and Zhi-Yao Yang prepared all the tables and figures. Cheng-Min Zhang proposed a theoretical model and reasoned in detail, analyzed the conclusions and put forward relevant suggestions. De-Hua Wang, Erbil Gügercinoğlu, Xiang-Han Cui,  Jian-Wei Zhang, Shu Ma, and Yun-Gang Zhou edited, wrote, and revised the paper. All authors reviewed the manuscript, and then discussed and revised the paper together, and finally agreed to submit the paper.

\section*{Funding}

This work is supported by the National Natural Science Foundation of China (grant Nos. 12163001, 12463007 and U1938117), and the New Academic Seedling Foundation of
Guizhou Normal University (grant No. [2022]05). EG is supported by National Natural Science Foundation of China (NSFC) programme 11988101 under the foreign talents grant QN2023061004L. Finally, we thank the anonymous referee for the valuable comments and suggestions, which have signiﬁcantly improved the quality of the paper.

\section*{Ethics Declarations}

{\bf Competing interests} The authors declare no competing interests.

\end{document}